\documentclass[usenatbib]{mn2e}

\usepackage{graphicx}
\usepackage{footmisc}
\usepackage{color} 
\usepackage{rotating}
\usepackage{amssymb}

\newcommand{\Msol}{M_\odot}

\newcommand{\kms}{{\rm km\,s^{-1}}}

\newcommand{\xmm}{\textit{XMM-Newton }}

\newcommand{\swift}{\textit{Swift }}

\newcommand{\integral}{\textit{INTEGRAL }}

\newcommand{\gmm}{$\Gamma$ }                                
\newcommand{\wgmm}{$\left<\Gamma\right>$ }                  
\newcommand{\sigwgmm}{$\sigma\sb{\left<\Gamma\right>}$ }    
\newcommand{\gmmpart}{$\Gamma\sb{\rm{part}}$ }              
\newcommand{\gmmfull}{$\Gamma\sb{\rm{full}}$ }              
\newcommand{\xspec}{\texttt{XSPEC} }
\newcommand{\chisq}{$\chi\sp{2}$ }
\newcommand{\lumx}{$L\sb{X}$ }                   
\newcommand{\hb}{H$\beta$ }                      

\title[X-ray spectra of type 1 AGN]{New constraints on the X-ray spectral properties of type 1 AGN}
\author[Scott et al.]{A.E. Scott$^{1}$\thanks{E-mail: aes25@star.le.ac.uk}, G.C. Stewart$^{1}$, S. Mateos$^{1,2}$, D.M. Alexander$^{3}$, S. Hutton$^{3}$ and M.J. Ward$^{3}$ \\
$^{1}$Department of Physics and Astronomy, University of Leicester, University Road, Leicester LE1 7RH, UK\\
$^{2}$Instituto de Fisica de Cantabria (CSIC-UC), Avenida de los Castros, 39005 Santander, Spain \\
$^{3}$Department of Physics, Durham University, South Road, Durham, DH1 3LE, UK}

\begin{document}

\date{Accepted YYYY Month DD. Received YYYY Month DD; in original form YYYY Month DD}

\pagerange{\pageref{firstpage}--\pageref{lastpage}} \pubyear{2011}

\maketitle

\label{firstpage}

\begin{abstract}
We present a detailed characterization of the X-ray spectral properties of 761 type 1 AGN, selected from a cross-correlation of the SDSS DR5 quasar catalogue and the incremental version of the second \xmm serendipitous X-ray source catalogue 2XMMi-DR2.  The X-ray spectrum of each source is fit with models based on a simple power law to which additional cold absorption and/or soft excess features are added if an F-test at 99\% significance requires them.  The distribution of best-fitting photon indices, $\Gamma$, is fit with a Gaussian with mean \wgmm = $1.99 \pm 0.01$ and dispersion \sigwgmm = $0.30 \pm 0.01$, however this does not provide a good representation of the distribution due to sources with very flat or steep \gmm values.  A highly significant trend for decreasing \gmm values with increasing 2-10 keV luminosity, $L\sb{X}$, is seen but only a weak trend with redshift is found.  Intrinsic cold absorption is detected in $\sim 4\%$ of the sample and soft excess emission is detected in $\sim 8\%$.  These values are lower limits due to the detectability being limited by the quality of the spectra and we suggest the intrinsic values may be as high as $\sim 25\%$ and $\sim 80\%$ respectively.  The levels of rest-frame absorption are higher than expected for type 1 objects ($N\sb{H} = 10\sp{21}-10\sp{23} \textrm{ cm}\sp{-2}$) and the fraction of absorbed sources and the $N\sb{H}$ values are not seen to vary with \lumx or $z$.  The average blackbody temperature used to model the soft excesses is $\left<\textrm{kT}\right> = 0.17 \pm 0.09$ keV.  This temperature is found to correlate with \lumx but not the blackbody luminosity or the black hole mass which do correlate with each other.  A strong correlation is found between the luminosities in the blackbody and power law components suggesting that a similar fraction is re-processed from the blackbody to the power law component for the entire luminosity range of objects.  A positive correlation between \gmm and the X-ray derived Eddington ratio is found for the sources whose mass was determined using the \hb line, but a negative correlation is found if the CIV line is used.  No correlation is found if the MgII line is used.  No significant correlations are found between the blackbody temperature, luminosity or black hole mass with Eddington ratio, despite a link between the power law and blackbody production being indicated.  The sample includes 552 confirmed RQQ and 75 confirmed RLQ.  The RLQ are found to have higher \lumx values than their RQQ counterparts, suggesting an additional X-ray component, perhaps related to a jet, is present in these sources.  This component may also be the cause of the flatter \gmm values seen in RLQ.
\end{abstract}


\section{Introduction}

It is widely accepted that Active Galactic Nuclei (AGN) are accreting supermassive ($M>10\sp{6}\Msol$) black holes (SMBH) residing at the centre of galaxies \citep{lynden-bell}.  They have bolometric luminosities which can be as high as $\sim 10^{48}$ erg s$^{-1}$ (e.g. \citealt{woo02}), with the X-ray emission (accounting for typically $\sim$10\% of the total luminous output) thought to originate from the innermost regions of the AGN. By studying their X-ray spectra we hope to further our understanding of the physical conditions surrounding the black hole.

The earliest observations of AGN revealed that their X-ray spectra were remarkably similar, and could be described by a simple power law \citep{mushotzky80}.  This primary X-ray emission is believed to be due to inverse Compton scattering of low energy UV photons from the accretion disc.  These photons are up-scattered to X-ray energies by interaction with the corona, a cloud of relativistic electrons found above the surface of the accretion disc \citep{haardt93}.  In higher quality X-ray spectra, additional features can be distinguished.  Below 1 keV, excess emission with respect to an extrapolation of the power law is sometimes detected \citep{arnaud85,turner89}, whose origin remains unknown.  The standard explanation for this `soft excess' is that it is thermal emission from the inner accretion disc and hence is usually modelled as a blackbody component.  The high energy tail of thermal emission from a standard thin accretion disc around a SMBH accreting at the Eddington rate is expected to have a blackbody temperature of $\sim 0.02-0.04$ keV \citep{shakurasunyaev73}, however the measured blackbody temperatures are typically higher than this (e.g. \citealt{gierlinskidone04}).  Lower energies are also affected by photoelectric absorption.  In some sources absorption levels are detected above that which is expected due to the Galactic component.  This may be due to cold or ionized (warm) gas intrinsic to the AGN, or due to its host galaxy.  At higher energies ($10-300$ keV), a hump may also be detected due to the Compton reflection of the incident power law continuum photons off the accretion disc. Despite being detected in the X-ray spectra of Seyferts, the evidence for this component in quasars is poor, since their emission is largely dominated by the luminous power law continuum.  It has been suggested that the reflection component could also be intrinsically weaker in high luminosity quasars since the inner accretion disc is more likely to become ionized, thus reducing the amount of neutral matter remaining, which is required to create a reflection feature (e.g. \citealt{mushotzky93})

The power law emission can be described by  P$\sb{E} \propto E\sp{-\Gamma}$, where the spectral (photon) index, $\Gamma$, was originally found to be  approximately 1.7 \citep{turner89}.  As higher quality spectra became available the more complex features, including reflection, were discovered in addition to this underlying power law \citep{pounds90}.  When the reflection component is modelled separately, the spectral index of the power law is adjusted to \gmm $\sim 1.9$ \citep{nandra94}.  Many subsequent studies have attempted to constrain the intrinsic value of this photon index, and determine how varied this value is between different objects.  

In general, most X-ray studies of AGN find an average \gmm value of $1.9-2.0$ \citep{corral11,mateos10,young09,mainieri07,mateos05b,mateos05a,page06,just07,green09,mittaz99} and an intrinsic dispersion of $\sigma \sim 0.2-0.3$ \citep{mateos10,young09,mainieri07,tozzi06,mateos05b,dai04,reeves00}.  However, different average values are reported for samples created by different satellites because of the selection effects occurring due to the different band passes.  For example, samples selected in hard X-rays by \integral or \swift yield generally lower average spectral indices of $\sim 1.7$ \citep{molina09, winter09}, likely because there is a bias towards detecting flatter sources at higher X-ray energies.  The \gmm value reported is dependent upon the energy range it is considered over, particularly if some spectral components are not properly modelled.  The presence of a warm absorber or unmodelled reflection component can produce a slope flatter than the true intrinsic value.  \citet{bianchi09a} suggest that the difference in photon index found between different populations of AGN could be due to varying amounts of Compton reflection present in the sources.  For their sample, quasars were found to have steeper power law slopes ($\Gamma = 1.80 \pm 0.05$) than Seyferts ($\Gamma = 1.66 \pm 0.05$), which may be indicative of a larger reflection component present in the Seyfert population.  When a reflection component was included in the models, the \gmm values of both populations increased, but more so for the Seyferts.

Different \gmm values in other populations of AGN are likely due to real physical differences.  For example, radio loud quasars (RLQ) are known to have flatter power law slopes than their radio quiet counterparts (RQQ) (e.g. \citet{reeves00} RLQ: $\Gamma = 1.66 \pm 0.04$, RQQ: $\Gamma = 1.89 \pm 0.05$), likely due to their X-ray spectra being contaminated by synchrotron emission from their powerful relativistic radio jets \citep{sambruna99}.  Narrow Line Seyfert 1s (NLS1) are also observed to have steeper power law slopes than their broad-line counterparts (BLS1) (e.g. \citet{brandt97} NLS1: $\Gamma = 2.15 \pm 0.05$, BLS1: $\Gamma = 1.87 \pm 0.04$), which could be due to these objects accreting at high rates. For higher mass accretion rates, a stronger soft X-ray/UV component from the disc is created, which in turn causes more Compton cooling, producing a steeper X-ray spectrum.  This scenario was offered as an explanation for the NLS1 RE 1034+39, which has an unusually soft power law of \gmm $\sim$ 2.6 \citep{pounds95}.  The spectral state observed was noted to be similar to the high/soft state observed in galactic black hole binaries that are accreting close to the Eddington limit \citep{remillard06}.

Despite being widely investigated, it is still not clear whether there is any intrinsic variation of \gmm with X-ray luminosity or redshift.  It is natural to assume that the X-ray production is related to the black hole mass, accretion rate and luminosity, however this dependence is poorly understood. \citet{dai04} first reported a correlation with luminosity that was positive, implying that the X-ray spectral slopes were steeper in higher luminosity sources, however later studies either report no correlation \citep{winter09,tozzi06,mateos05b,mateos05a,shemmer05,reeves00} or a negative correlation \citep{corral11, young09,saez08,reeves97} implying that the X-ray spectral slope flattens as the X-ray luminosity of the sources increases.  If a trend with redshift was found it may indicate that the accretion mechanism was different for objects in the past.  Many studies of large samples of AGN report no evolution of \gmm with redshift \citep{young09,mainieri07,just07,tozzi06,page06,mateos05b,mateos05a,perola04,piconcelli03,reeves00,green09} which agrees with some results of studies specifically targeting a few high redshift objects and measuring \gmm values consistent with those for lower redshift objects \citep{vignali05,shemmer05}.  However, some report \gmm flattening for higher redshift sources \citep{kelly07,bechtold03,vignali99}.


The Unified Model of AGN \citep{antonucci93} states that different classes of AGN are examples of the same physical object but viewed at different angles.  In type 1 objects we are able to observe the inner most region, whereas in type 2 objects this is shielded from view by a dusty torus.  Therefore type 1 AGN are expected to be unabsorbed in X-rays and hence require no intrinsic absorption components in the modelling of their spectra.  However, previous studies have found both type 2 objects which appear to be X-ray unabsorbed \citep{panessa02} and samples of type 1 AGN with significant fractions of absorbed sources \citep{mateos10,young09,mateos05b}.  These results are not reconciled by the standard orientation based Unified Model.

Soft excess components are found to be common features in the X-ray spectra of type 1 AGN.  Originally thought to be present in $30\%$ of hard X-ray selected Seyferts \citep{turner89}, more recent estimates give $\sim 7\%$ (e.g. \citealt{mateos10}).  However the origin of this component is still a matter of debate.  It was originally interpreted as the hard tail of the `Big Blue Bump' seen in the UV due to the accretion disc, but is now thought to be either an artefact of ionized absorption \citep{gierlinskidone04} or ionized reflection \citep{ross05,crummy06}.

One of the key questions in AGN research is to explain why different classes of AGN have different spectral properties.  The best way to do this is though a statistical study of a large number of AGN.  \citet{bianchi09a} investigate the X-ray spectral properties of $\sim 160$ AGN from targeted \xmm observations.  This sample has the advantage of high signal-to-noise (S/N) X-ray spectra, however it is also biased towards sources at low redshift (90\% of their sources are at $z<1$).  Since \xmm targets have been used, the sample contains objects observed for a number of different reasons, and hence does not represent a uniform sample of AGN.  \citet{mateos10} presented an X-ray spectral analysis of $\sim 500$ type 1 AGN from the \xmm Wide Angle Survey (XWAS).  The overall broad-band properties and their dependences on X-ray luminosity and redshift were investigated, offering a complete picture of the X-ray spectral properties of an X-ray selected sample of type 1 AGN.  \citet{young09} present an optically selected sample of $\sim 500$ objects by cross-correlating the Sloan Digital Sky Survey (SDSS; \citealt{SDSS}) and the \xmm archive.  They fit only simple models to their spectra and include only $\sim 300$ of their sources in their investigation of \gmm with $L\sb{X}$.  It is possible that by restricting their sample to the higher S/N sources, they have introduced a bias towards lower redshift objects, or the most X-ray luminous higher redshift sources.   

In this study we analyze the X-ray spectra of the largest sample of type 1 AGN to date ($\sim 750$).  They are optically selected from SDSS and also have observations taken by \textit{XMM-Newton}.  We fit a range of spectral models to each of our sources, over an energy range of 0.5-12.0 keV, including ones to model any soft excess components in order to gain insight into their nature.  We choose the best-fitting model for each source and constrain the average power law slope and intrinsic dispersion.  We also investigate any absorption found in the sources.  Our sample covers a large range in redshift, with sources up to $z \sim 5$, and covers around 4 orders of magnitude in $L\sb{X}$.  This allows us to investigate the dependencies of \gmm with X-ray luminosity, $L\sb{X}$, and redshift, \textit{z}. 

This paper is organized as follows.  In Section~\ref{section:procedure} we describe the properties of the two initial catalogues, DR5QSO and 2XMMi and the extraction of the X-ray spectra and the subsequent spectral fitting.  Sections~\ref{section:gmm}, \ref{section:gmmz} and \ref{section:gmmlx} consider the overall distribution of power law slope values and the correlations of \gmm with redshift and luminosity.  Section \ref{section:radio} explores the radio properties of the sample.  Section~\ref{section:apo} considers the properties of the sources which required an absorbed power law model, whilst those requiring a soft excess component are discussed in Section~\ref{section:po+bb}.  In Section~\ref{section:black_hole} we use black hole mass estimates from the literature to investigate how the accretion properties affect the observed X-ray emission.  A discussion of our results and conclusions are presented in Sections~\ref{section:discussion} and \ref{section:conclusion} respectively.

In this paper we assume a cosmology from the WMAP results where H$\sb{0}$=70 km s$\sp{-1}$ Mpc$\sp{-1}$, $\Omega\sb{M}$ = 0.3 and $\Omega\sb{\Lambda}$ = 0.7 \citep{spergel03}.


\section{Procedure}
        \label{section:procedure}
        \subsection{Data Sample}

The sample of objects for this study was created by a positional cross-correlation of the SDSS DR5 quasar catalogue \citep{DR5QSO} and 2XMMi \citep{2XMM}, using a 10\arcsec\ matching radius.  This initial sample consisted of 2013 matches, corresponding to 1281 unique objects.

The DR5 quasar catalogue is produced from the fifth data release of SDSS \citep{DR5}, which provides imaging in 5 broad optical bands (\textit{ugriz}).  Quasar candidates are primarily selected for spectroscopic follow up based on their location in multidimensional SDSS colour space, however others are selected if they lie within 2.0\arcsec\ of a FIRST (Faint Images of the Radio Sky at Twenty-Centimeters; \citealt{FIRST}) radio source, introducing a possible radio bias into the catalogue. 6.6\% of our final sample of sources were selected for follow up due to a close match with a FIRST detection, however all but 2 of these sources would have been included automatically by the colour selection. The catalogue includes objects with an absolute \textit{i} band magnitude brighter than $-22$, an apparent \textit{i} band PSF magnitude fainter than 15, a reliable redshift and at least one emission line in their spectra with a FWHM $>1000~\kms$.  77,429 quasars are included with redshifts from 0.08 to 5.41, covering a sky area of about 5740 deg$^2$. 

2XMMi \citep{2XMM} is the incremental second serendipitous X-ray source catalogue, consisting of both the targets and non-target source detections that are recorded in the large (30\arcmin) field of view of \textit{XMM-Newton's} EPIC (European Photon Imaging Camera; \citealt{mos,pn}) X-ray cameras.  The catalogue contains information from 4117 pointed observations covering a non-overlapping sky area of $\sim$ 420 deg$^2$.  It contains $\sim 290,000$ detections, which due to field duplication by multiple observations, corresponds to $\sim 220,000$ unique sources.  Since observations with \xmm can be requested for a variety of reasons, including the target detections in our sample could add bias towards `interesting' objects.  Our final sample includes 62 sources which were the target of an observation (8\% of the sample), which are naturally the ones with the highest number of counts and the sources fit with the most complex models.  Since our analysis deals primarily with the distribution of \gmm values and trends of \gmm with various quantities, we investigate whether the target sources have a significantly different \gmm distribution compared with the sample of serendipitously detected sources.  A  Kolmogorov--Smirnov (KS) test finds them to be consistent with each other and therefore the use of the target detections in our data sample should not affect our overall findings.

We determine the radio loudness of the objects in our sample using the definition $R\sb{L}=F\sb{R}/F\sb{O}$ where $F\sb{R}$ is the flux at 5 GHz and $F\sb{O}$ is the flux at $4400\textrm{\AA}$.  Any sources for which $R\sb{L} > 10$ are taken to be radio loud \citep{kellermann89}.  To do this we cross-correlated our X-ray sample with the FIRST \citep{FIRST} catalogue which covers a similar area to SDSS at a frequency of 1.4GHz to a sensitivity of $\sim$ 1 mJy.  128 quasars had a match with a FIRST source, 44 fell outside the FIRST area, and for the remaining sources $5\times \textrm{RMS}$ value at the source position was used as an upper limit.  The flux measured by FIRST was extrapolated to a flux at 5 GHz by assuming the radio emission is described by a power law with spectral (energy) index $\alpha=-0.8$.  The flux at $4400\textrm{\AA}$ was determined from a linear interpolation of the optical fluxes in each of the SDSS \textit{ugriz} bands.  552 (72\%) were found to be radio quiet quasars, 75 (10\%) radio loud quasars, 90 (12\%) were sources whose upper limits are too high to determine their radio loudness with confidence and 44 (6\%) sources lay outside the FIRST coverage area and therefore a radio loudness could not be determined.  If all the sources with radio upper limits are assumed to be either all radio quiet or all radio loud, we find that the radio loud fraction lies between $\sim$10\% and $\sim$22\%, respectively,  in broad agreement with previously reported values (e.g. \citealt{kellermann89}).

        \subsection{Extraction of Spectral Products}
  
The X-ray spectra of all 2013 detections were extracted using an automatic pipeline incorporating tasks from the Science Analysis Software v6.6.0 (SAS\footnote{The description and documentation are available online at http://xmm2.esac.esa.int/sas/}), specifically designed for the reduction of \xmm data.  In particular the task \texttt{eregionanalyse} was used to select the extraction regions so as to optimize the S/N.  Circular regions with an average radius of  $\sim$ 17\arcsec were used for the source spectra.  A circular background region with a radius of 50\arcsec was automatically chosen in the same CCD and  close to the source, but excluding any other nearby sources.  Auxiliary response files and redistribution matrices were generated using \texttt{arfgen} and \texttt{rmfgen}, respectively.

Each detection was subjected to a manual screening process in which any that were severely affected by Out of Time events or contamination from a nearby bright source were removed from the sample.  Sources falling into chip gaps or partially off the edge of the field of view were also removed.  Multiple detections of the same source were merged into a single spectrum, with the data from the two MOS cameras being combined, but a separate MOS and pn spectrum for each source generated, assuming that data from both detectors was available.  For 278 sources data from only one detector was available or hadn't been rejected in the screening process.  For these sources, spectral fitting was carried out on the single spectrum available.  6\% of the original sample of unique objects were removed in the screening process, leaving a sample of 1201 AGN.  Objects with $<75$ total counts were excluded, leaving 761 sources.

        \subsection{Spectral Fitting}
	\label{section:fits}

The spectra were binned using the \texttt{grppha} Ftool\footnote{http://heasarc.gsfc.nasa.gov/docs/software/ftools/\label{ftool}}, to ensure that Chi squared ($\chi\sp{2}$) statistics could be used. The minimum number of counts per bin was varied depending on the S/N of the spectra, with 15 being used for sources with $\sim 100$ counts, increasing up to 30 counts per bin for sources with a few thousand counts.  The spectra were then fit with physically motivated models using \xspec v11.3.2 \citep{xspec} over the energy range 0.5-12.0 keV.  Where available, the MOS and pn data were fit simultaneously with the same parameters and a freely varying constant was added to the model to account for calibration offsets between the two cameras \citep{mateos09}.  Each model included an absorption component fixed at the value of the Galactic column density in the direction of the source, as determined from the \texttt{nH} Ftool\footref{ftool} which uses the HI map of \citet{dickey90}.

The 761 sources with more than 75 counts were fit with both a simple power law (`po') and an absorbed power law (`apo'), in which \gmm and an intrinsic $N\sb{H}$ component at the redshift of the source were allowed to vary freely.  For 680 sources with more than 100 counts, models including a blackbody component were also considered.  This component was introduced to model any soft excess, and was added to both the simple power law and absorbed power law models.  A summary of the different models considered, along with their \xspec terminology can be found in Table~\ref{table:models}.  Absorption was modelled using \texttt{phabs} with abundances from \citet{wilms00} and cross-sections from \citet{cross}.

\begin{table*}
\begin{minipage}{170mm}
\centering
  \caption{This describes the different models used to fit the X-ray spectra of the sources.  The spectral components included in each particular model are listed in column 2.  Galactic absorption is also included in each, modelled by the additional \texttt{phabs} component.  The 4th column lists the total number of sources in the sample which were best fit by that particular model as chosen by an F-test at 99\% significance.  The final column lists the number of those sources which are also \textit{well fit} by the best-fitting model i.e. H$0>1\%$.  We note that the blackbody component is only considered in spectra with $>100$ counts (see Section~\ref{section:po+bb}\label{bb})}
  \label{table:models}
     \centering
     \begin{tabular}{lllccl}
     \hline
     Model & Components & \xspec terminology & No. of sources & \multicolumn{2}{c}{No. of good fits} \\
      &  &  &  & \multicolumn{2}{c}{H$0>1\%$}  \\
     \hline
     po                    & Power Law                                & phabs*po               &  672 &           650 &  (89\%) \\
     apo                   & Power Law + intrinsic absorption                & phabs*zpha*po          &  29\footnote{1 source shows a Fe K$\alpha$ line and is fit with the model pha*(zpha*po+zgauss)}&         28 &  (4\%) \\
     po+bb                 & Power Law + soft excess                         & phabs*(zbbody+po)      &  55 &            52 &  (7\%)  \\
     apo+bb                & Power Law + intrinsic absorption + soft excess  & phabs*(zbbody+zpha*po) &  5\footnote{2 sources show Fe K$\alpha$ lines and are fit with the model pha*(zgauss+zbbody+zpha*po)}&       4 & ($<$1\%) \\
     \hline
                           &                                          &                              & 761            & 734   \\
     \hline
  \end{tabular}
\end{minipage}
\end{table*}

\begin{figure*}
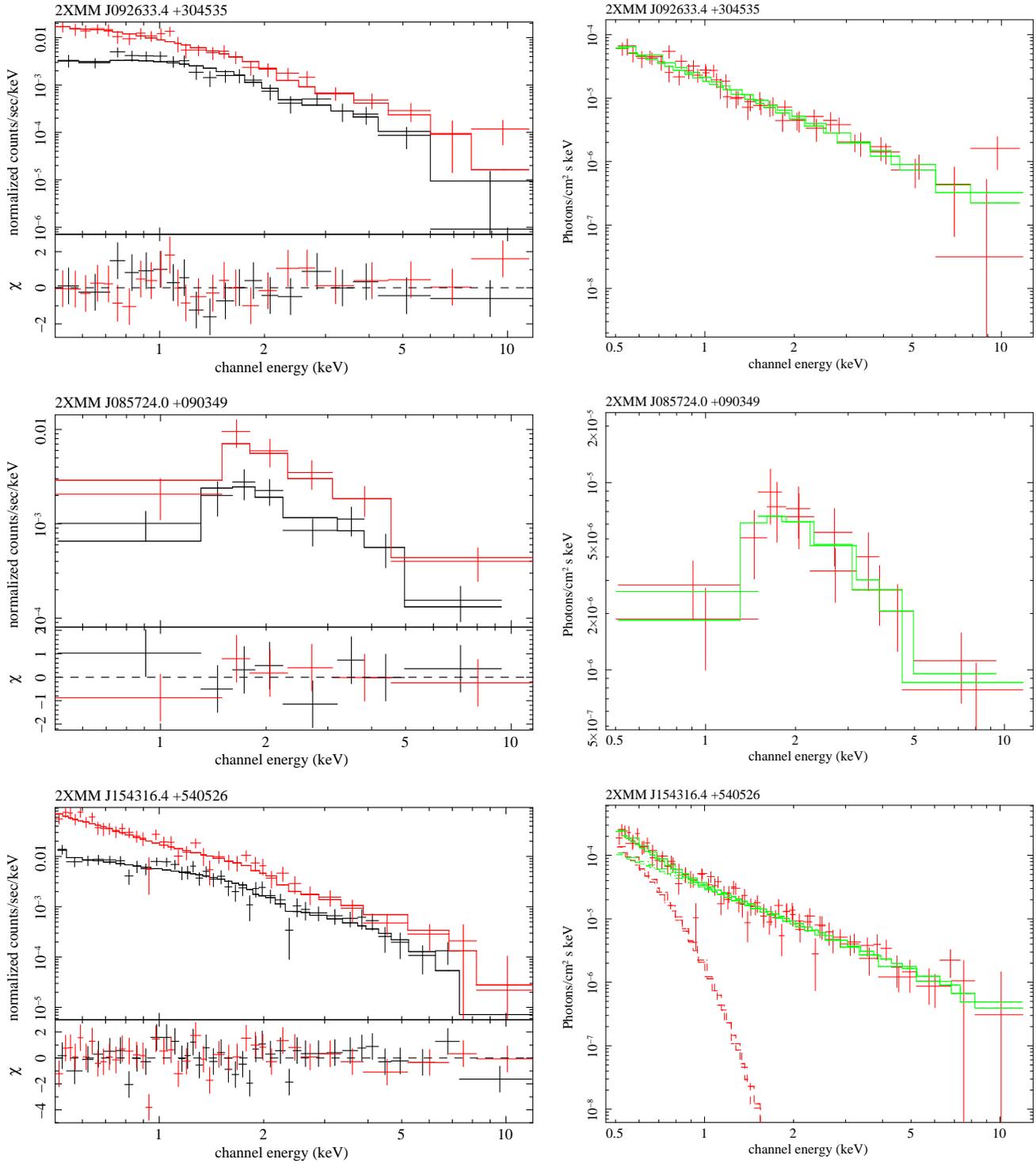

  \centering
  \begin{tabular}{cc}
       \includegraphics[angle=270,width=0.5\textwidth]{paper_75084_data.ps}&
       \includegraphics[angle=270,width=0.45\textwidth]{paper_75084_ufspec.ps}\\
       \includegraphics[angle=270,width=0.5\textwidth]{paper_71597_data.ps}&
       \includegraphics[angle=270,width=0.45\textwidth]{paper_71597_ufspec.ps}\\
       \includegraphics[angle=270,width=0.5\textwidth]{paper_141491_data.ps}&
       \includegraphics[angle=270,width=0.45\textwidth]{paper_141491_ufspec.ps}\\
  \end{tabular}
  \caption{This figure includes the raw (left) and unfolded (right) spectra for 3 of the sources in the sample which are each best fit with different models.  2XMM J092633.4+304535 (top row) is best fit with the simple power law model (po), where \gmm$=1.98 \pm 0.07$. 2XMM J085724.0+090349 (middle row) is best fit with the absorbed power law model (apo), where \gmm$=1.85 \pm 0.36$ and $N\sb{H}=(9.8 \pm 3.6) \times 10\sp{22} \textrm{ cm}\sp{-2}$.  2XMM J154316.4+540526 (bottom row) is best fit with a power law model and a soft excess (po+bb), where \gmm$=1.90 \pm 0.09$ and $\textrm{kT}=0.10 \pm 0.02$ keV.  The sources include (from top to bottom) 630, 129 and 1131 total (MOS+pn) counts.  In the left panels, the MOS data is shown in black and the pn in red.  In the bottom right panel, the dashed line indicates the contribution of the blackbody component to the overall spectral model.}
  \label{fig:spectra}
\end{figure*}

The best-fitting model for each source was chosen via an F-test at 99\% significance.  The best fit was assumed to be a simple power law unless an F-test determined that an additional intrinsic absorption component or blackbody component was required.  These sources now assumed to be best fit with the `apo' or `po+bb' models were then tested to see if the best fitting model was in fact `apo+bb' i.e. both components were required.  The number of sources best fit with each model can also be found in Table~\ref{table:models}.  Example spectra of sources best-fit with different models are shown in Figure~\ref{fig:spectra}.

\begin{figure}
  \centering
  \includegraphics[width=0.47\textwidth]{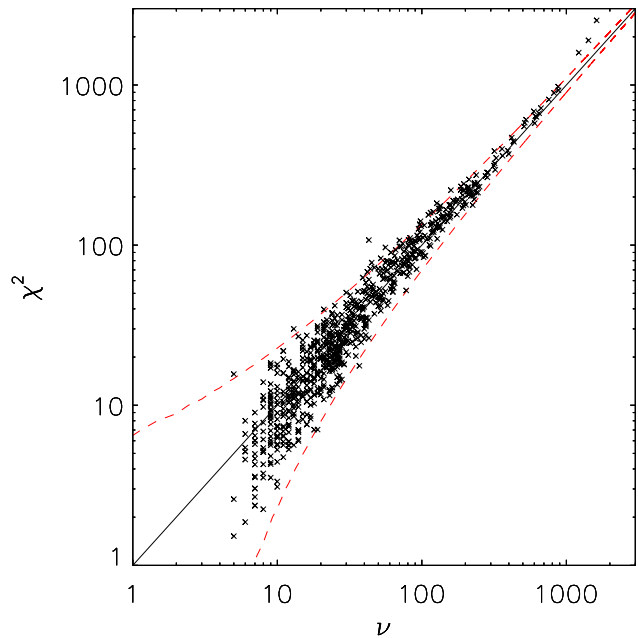}  
  \caption{Plot of Chi squared value ($\chi\sp{2}$) vs Degrees of freedom ($\nu$) for each of the X-ray spectra.  The solid line indicates the location of $\chi\sp{2}/{\nu} = 1$.  The dashed red lines represent the locus outside which we would expect to observe \chisq values only 1\% of the time if the fit is correct.}
  \label{fig:chi}
\end{figure}

While the F-test determined the \textit{best fit} to the data, a model was deemed to be a \textit{good fit} to the data if the null hypothesis probability (H0) was $>1\%$.  The majority (96\%) of the sample were well fit by the model chosen by the F-test and the numbers of good fits for each type of model can be seen in Table~\ref{table:models}.  For the sources with high numbers of counts (i.e. $>1000$ degrees of freedom) the models we fit are not sufficient to deal with the complexities shown to exist from studies of bright, nearby AGN (e.g. \citealt{pounds04}), and therefore $\textrm{H0}<1$\%.  For a good fit we expect $\chi\sp{2}/{\nu} \sim 1$, which is shown in Figure~\ref{fig:chi} and compared to the actual quality of the fits.  The region in which the best fit model is considered correct at the 99\% confidence level is shown between the red dashed lines.

The spectral fits of each source, particularly those with a poor original fit, were visually inspected and revealed 3 sources with evidence for the Fe K$\alpha$ emission line.  A Gaussian component, \texttt{zgauss}, was manually added to the model for these sources. We note that evidence for this component likely exists in more of the source spectra, although a systematic search for this component is beyond the scope of this paper.  The 3 Gaussian components have energies (E) and equivalent widths ($\sigma$) as follows: $\textrm{E}=6.72 \pm 0.002$ keV and $\sigma = 0.07 \pm 0.01$ keV, $\textrm{E}=6.67 \pm 0.03$ keV and $\sigma = 0.22 \pm 0.03$ keV, $\textrm{E}=6.40 \pm 0.05$ keV and $\sigma = 0.13 \pm 0.02$ keV, which are typical of those found for nearby, bright AGN. 

\begin{figure}
  \centering
     \includegraphics[width=0.45\textwidth]{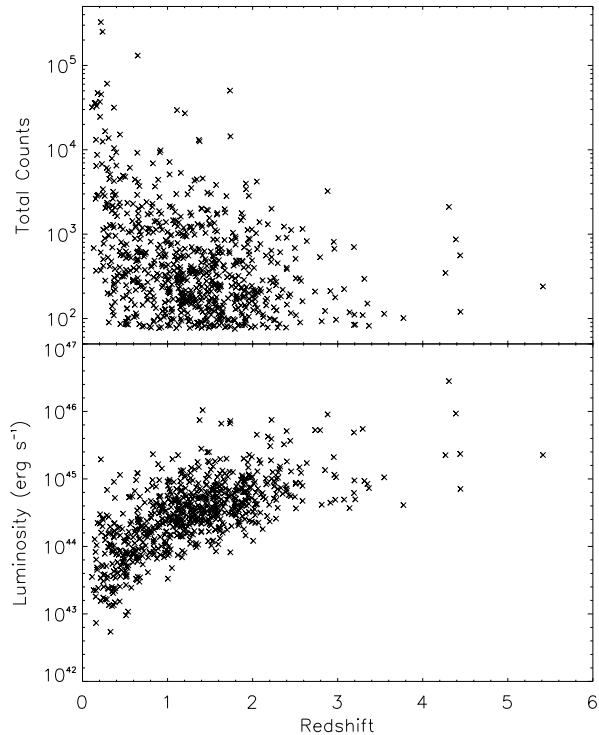}
     \vspace{2mm}
  \caption{Plots showing the general properties of the final sample of 761 objects with X-ray spectral fits.  Top - The redshift and counts distribution.  Bottom - The luminosity (2-10 keV) and redshift range of the sample.}
  \label{fig:properties}
\end{figure}

Fluxes and rest frame luminosities for the sources are computed from the best fit model which were corrected for Galactic absorption, and in the case of the luminosities, intrinsic absorption where necessary.  The final sample covers a large range in redshift (0.11 to 5.41) and 2-10 keV X-ray luminosity ($10\sp{43} - 10\sp{46} \textrm{ erg s}\sp{-1}$), better enabling us to investigate the dependencies of \gmm on these quantities.  The redshift and counts distribution can be seen in Figure~\ref{fig:properties} (top) and the luminosity-redshift dependence can be seen in Figure~\ref{fig:properties} (bottom).  The properties of our sample are consistent with those of the entire DR5 quasar catalogue in terms of both the average redshift and average absolute \textit{i} band magnitude.  More details will be presented in Hutton et al. (in prep).

\begin{figure}
  \centering
  \includegraphics[width=0.48\textwidth]{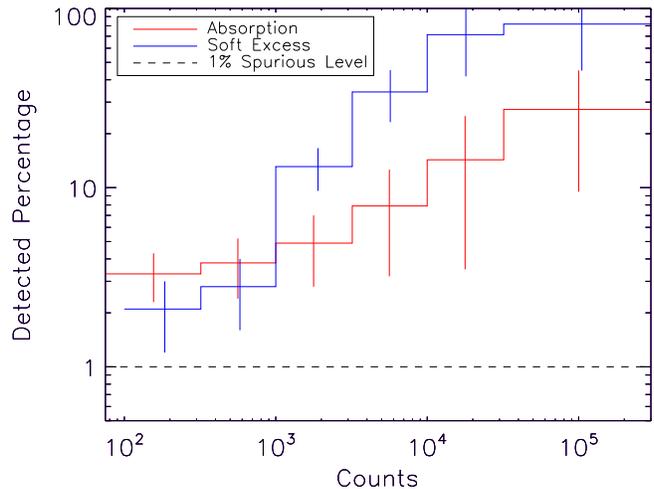}  
  \caption{This plot shows how the percentage of sources that are detected with an absorption or soft excess component varies depending on the number of counts that are available in the spectra.}
  \label{fig:detection}
\end{figure}

As expected, the majority of our sources, $\sim 90\%$, are best fit with the simple power law model and do not require any additional spectral components.  Figure~\ref{fig:detection} shows the fraction of sources that are detected with either an absorption component or a soft excess component as a function of the number of counts in the spectra.  At the highest count levels approximately 25\% of the sources require an absorption component while over 80\% require a soft excess. Both fractions fall with decreasing counts. As this could be due to our reducing sensitivity to these extra components we have performed simulations of the detectability of typical values of these as a function of counts (Scott et al. in prep). We find that these simulations reproduce almost exactly the behaviour observed in Figure~\ref{fig:detection}, suggesting that the level of required soft excess and absorption within the overall type 1 AGN population could be as high as the 80\% and 25\% found in the highest count range.

We have also conducted a joint spectral fitting of sources in narrow redshift bins to see if the additional spectral features are detected in a combination of lower count ($<500$) spectra which did not previously require them in the fit (i.e. were best-fit with the simple power law model).  We did this by fixing \gmm$=2.0$ for each source and fit the apo (or po+bb) model to the spectra, to find a single $N\sb{H}$ (or kT) value for all objects.  We expect from Figure~\ref{fig:detection}, that fitting spectra with $\sim 7000$ combined counts, will reveal the soft excess feature, as it should be detected $25-40\%$ of the time.  We find that for the lower redshift bins ($z<1$), the addition of a soft excess component, does result in a better fit to the combined spectra and the component has a temperature consistent with those reported in Section~\ref{section:po+bb}.  At $\sim 7000$ counts, we only expect $3-10\%$ sources to be detected with an absorption component.  Therefore in our sub-bins of $\sim 30$ sources, we do not expect to recover a significant absorption component.  This is what we find, with no significant improvement of the fit upon the addition of an intrinsic $N\sb{H}$ component.  Further details of this analysis will be presented in Scott et al. (in prep).

\section{Results}
  
   \subsection{Distribution of Power Law Slopes}
   \label{section:gmm}

\begin{figure}
  \centering
  \includegraphics[width=0.5\textwidth]{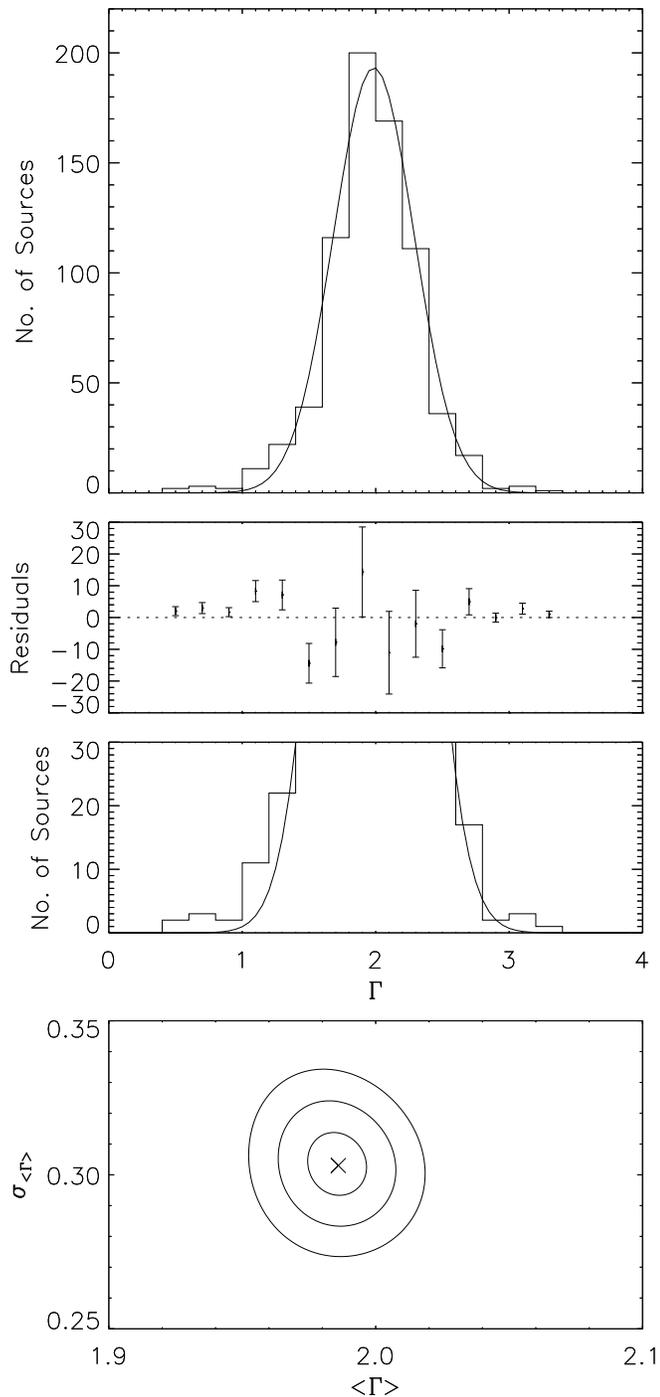}
  \caption{The histogram (top) shows the distribution of best-fitting photon indices where the required model gives an acceptable fit to the data.  The smooth line is the best-fitting Gaussian distribution, as determined by \chisq minimization and being described by \wgmm and $\sigma\sb{\left<\Gamma\right>}$.  Also included is a plot of the residuals, comparing the data to the model, and an enlarged version of the lower regions of the histogram to better highlight the presence of the extreme sources.  The bottom figure shows the best-fitting values, indicated by the cross, and the 1$\sigma$, 2$\sigma$ and 3$\sigma$ confidence contours for the derived values of \wgmm and $\sigma\sb{\left<\Gamma\right>}$.}
  \label{fig:gamma_dist}
\end{figure}

The distribution of best-fitting power law slope values, $\Gamma$, can be seen in Figure~\ref{fig:gamma_dist}.  It shows that most sources have a power law slope value $\sim 2$, with the arithmetic average value being \gmm$=1.97 \pm 0.01$.  The typical (68\%) error on a \gmm measurement is $\Delta\Gamma = 0.13$, significantly smaller than the spread of \gmm values, indicating that the observed dispersion of values is truly intrinsic.  The panel showing the enlarged distribution indicates that there are some sources with extremely flat ($\Gamma < 1.4$) or steep ($\Gamma > 2.6$) power law slopes.

\begin{figure}
  \centering
  \includegraphics[width=0.48\textwidth]{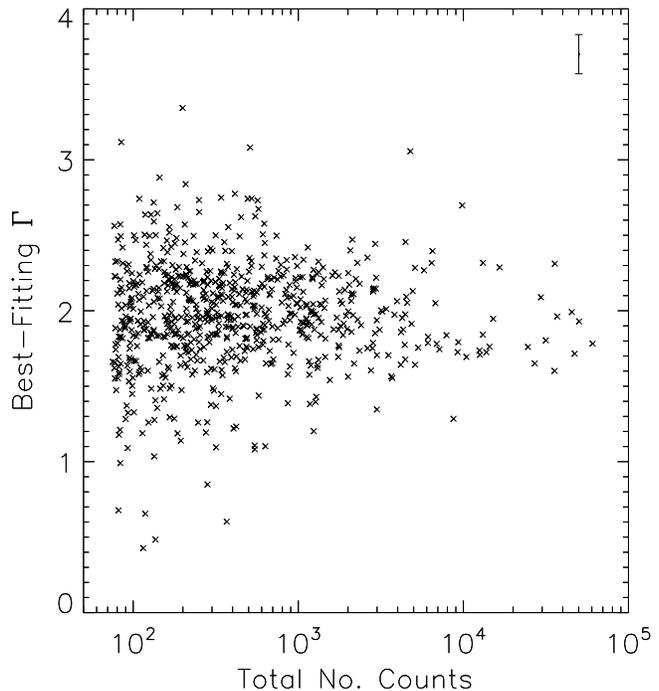}
  \caption{This shows the measured \gmm values against the total number of counts in each of the X-ray spectra for which $\textrm{H}0>1$\%.  The error bars have been omitted from this plot for reasons of clarity but an error of typical size can be seen in the top right hand corner.}
  \label{fig:snr}
\end{figure}

Such extreme values may be due to sources having spectra with low numbers of counts and hence the quality of the fit is poor.  Figure~\ref{fig:snr} shows the dependence of the derived \gmm value on the quality of the spectra, and it can be seen that many of the extremely flat or steep sources are indeed those with lower quality spectra.  This is likely due to the presence of intrinsic absorption or a soft excess component which are not detected in the spectra with enough significance.  However, we note that there are sources with high counts for which we do not believe this to be the case and therefore do have intrinsically flat or steep \gmm values.  A more detailed discussion of these sources is beyond the scope of this paper.  

We assume that we can represent the distribution of \gmm values with a Gaussian function of mean \wgmm and dispersion \sigwgmm and obtain simultaneous estimates of these values using a \chisq minimization technique.  Figure~\ref{fig:gamma_dist} (bottom) shows the 1$\sigma$, 2$\sigma$ and 3$\sigma$ confidence contours, along with the derived best-fitting values \wgmm = $1.99 \pm 0.01$ and \sigwgmm = $0.30 \pm 0.01$.  The best-fitting Gaussian can be seen overplotted on Figure~\ref{fig:gamma_dist} (top).  

\begin{figure}
  \centering
  \includegraphics[width=0.45\textwidth]{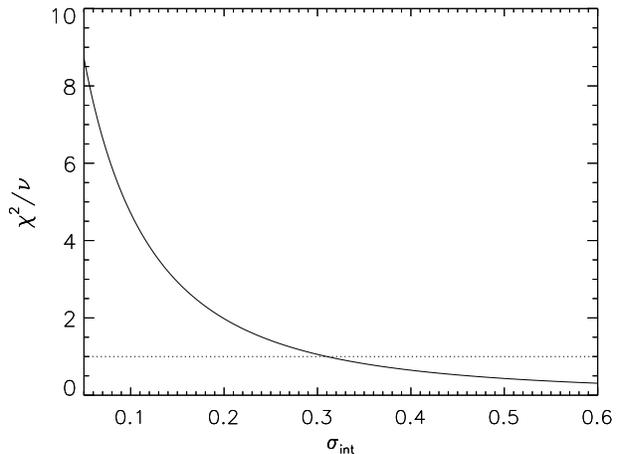}
  \caption{This shows how the reduced \chisq of the fit varies when increasing levels of intrinsic dispersion are added in quadrature to the individual errors on the \gmm values for each point in the sample.}
  \label{fig:dispersion}
\end{figure}

This analysis does not take into account the individual errors on each of the \gmm estimates, so here we investigate what contribution these make to the total dispersion we measure.  This is done by assuming that the observed dispersion can be expressed as the sum of the intrinsic dispersion and the individual errors on \gmm added in quadrature, i.e. $\sigma\sp{2} = \sigma\sp{2}\sb{int} + \sigma\sp{2}\sb{err}$.  We then calculate \chisq for a range of different values of $\sigma\sb{int}$, assuming the expected \gmm value for each point to be equal to the average value for the entire sample and the error on each point to be the 68\% error derived from its spectral fit.  The results of these fits can be seen in Figure~\ref{fig:dispersion}.  As expected, we find a `poor' fit for low intrinsic dispersion, while the fit becomes `too good' for much higher levels.  The point at which reduced \chisq becomes equal to 1 is taken as the best fitting value and we use the null hypothesis probabilities to determine the 68\% confidence interval on this value.  We find $\sigma\sb{int} = 0.31 \pm 0.01$, which is the same result that we obtain from our single Gaussian fit above.  We can therefore say that the individual errors on \gmm contribute very little to the overall dispersion.  This is not entirely unexpected since the typical 68\% error on \gmm is $\Delta\Gamma \sim 0.13$, which will contribute much less than an intrinsic dispersion of 0.30 when added in quadrature.

Our values of \gmm agree with those found by previous studies, including the XWAS sample, \gmm$=1.96 \pm 0.02$, $\sigma=0.27\sp{+0.01}\sb{-0.02}$ \citep{mateos10} and the \xmm COSMOS sample, \gmm$=2.06 \pm 0.08$ \citep{mainieri07}.  \citet{young09}, using a subset of objects from the same data presented here, find \gmm$=1.91 \pm 0.08$ and that their typical 68\% errors, $\Delta\Gamma = 0.15$ result in an intrinsic dispersion of $\sigma = 0.37$.

\begin{figure}
  \centering
  \includegraphics[width=0.5\textwidth]{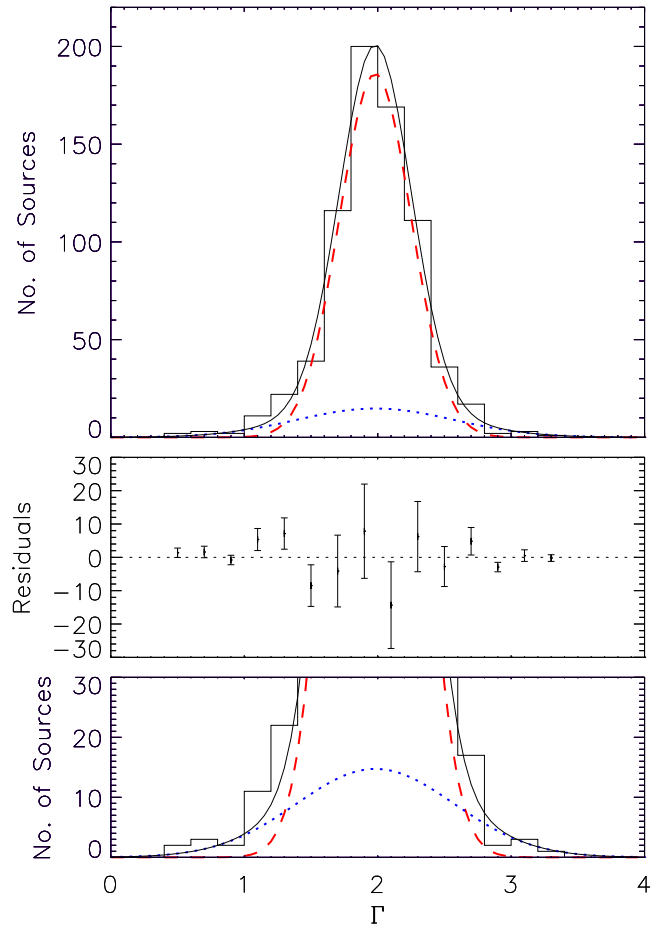}
  \caption{The red, dashed and blue, dotted lines show the two constituent Gaussians, while the black, solid line shows the overall model, produced by the addition of the two.}
  \label{fig:two_gauss}
\end{figure}

When fitting a single Gaussian to the \gmm distribution, a null hypothesis probability value of p=0.43\% was found.  Since this is $<1\%$, it is assumed that the model does not provide an acceptable fit to the data.  A two Gaussian model gives a better fit to the data (null hypothesis probability, p=15.4\%) and an F-test comparing the two Gaussian and one Gaussian models suggests that the inclusion of the second Gaussian is statistically valid (F-test probability of 99\%).  In this fit the average value of the two Gaussians, which were kept equal, was $\mu=1.98 \pm 0.01$, and the best-fitting dispersions of the Gaussians were found to be $\sigma\sb{1}=0.27 \pm 0.01$ and $\sigma\sb{2}=0.59 \pm 0.04$.  The fit can be seen in Figure~\ref{fig:two_gauss}.  We do not suggest that the 2 Gaussian model should be taken as a physically motivated fit to the \gmm distribution since it models both the flat and steep sources as a single `extreme' population, when in reality they are likely to be different sub-samples.  However since a multiple Gaussian model can produce an acceptable fit, where the single Gaussian is inadequate, it does highlight the existence of extreme sources in the distribution.  We test whether these are simply those with the lowest numbers of counts and find that a single Gaussian will provide an acceptable fit to the \gmm distribution only when sources with less than 300 counts are excluded, at which point the sample size is reduced to 385.  At this count limit, we find that $\sigma=3.7\times\Delta\Gamma$, where $\Delta\Gamma$ has fallen to 0.07, indicating that the dispersion is still significantly larger than the typical error.

   \subsection{Redshift Dependence of the Power Law Slope}
   \label{section:gmmz}

\begin{figure}
  \centering
  \includegraphics[width=0.48\textwidth]{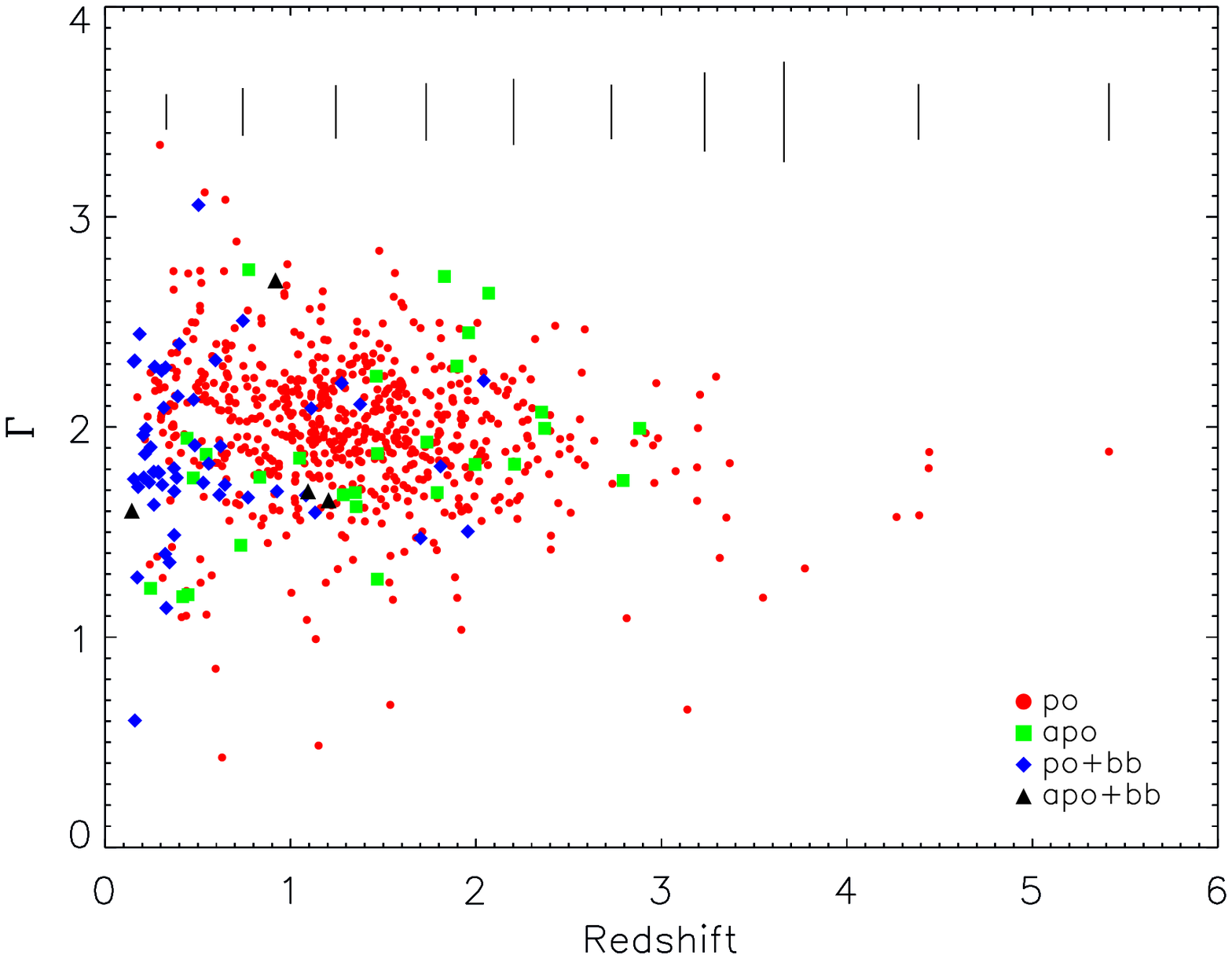}
  \caption{This plot includes the 734 sources with H0$>1\%$.  The different colours/symbols correspond to the best fitting model; red circles are po, green squares are apo, blue diamonds are po+bb and black triangles are apo+bb.  The error bars shown at the top of the plot correspond to the average of each of the $1\sigma$ errors on the individual sources, calculated in d$z$=0.5 bins.}
  \label{fig:gmm_z_scatter}
\end{figure}

In this Section we consider whether our data shows any evidence for an intrinsic variation of the spectral index, $\Gamma$, as a function of redshift.  This possibility has been widely investigated, but no clear consensus has been reached \citep{young09,mainieri07,just07,tozzi06,page06,mateos05b,mateos05a,perola04,piconcelli03,reeves00,green09,vignali05,shemmer05,kelly07,bechtold03,vignali99}.  Figure~\ref{fig:gmm_z_scatter} includes the 734 sources for which the best-fitting model provides an acceptable fit to the data i.e. H0 $>1\%$.  The sources are plotted with different colours/symbols according to their model.

We calculate Spearman's rank correlation coefficient, $\rho$, for both the entire population of 734 sources, and the subset of 650 sources whose best fit is a simple power law model (po).  Considering the entire population we find a $\sim3\sigma$ anticorrelation of \gmm with \textit{z} ($\rho = -0.11$, prob $= 0.002$).  Considering only the sources fit by po, we find a tighter $\sim5\sigma$ anticorrelation ($\rho$ = -0.20, prob = $4 \times 10\sp{-7}$).

\begin{figure}
  \centering
  \includegraphics[width=0.5\textwidth]{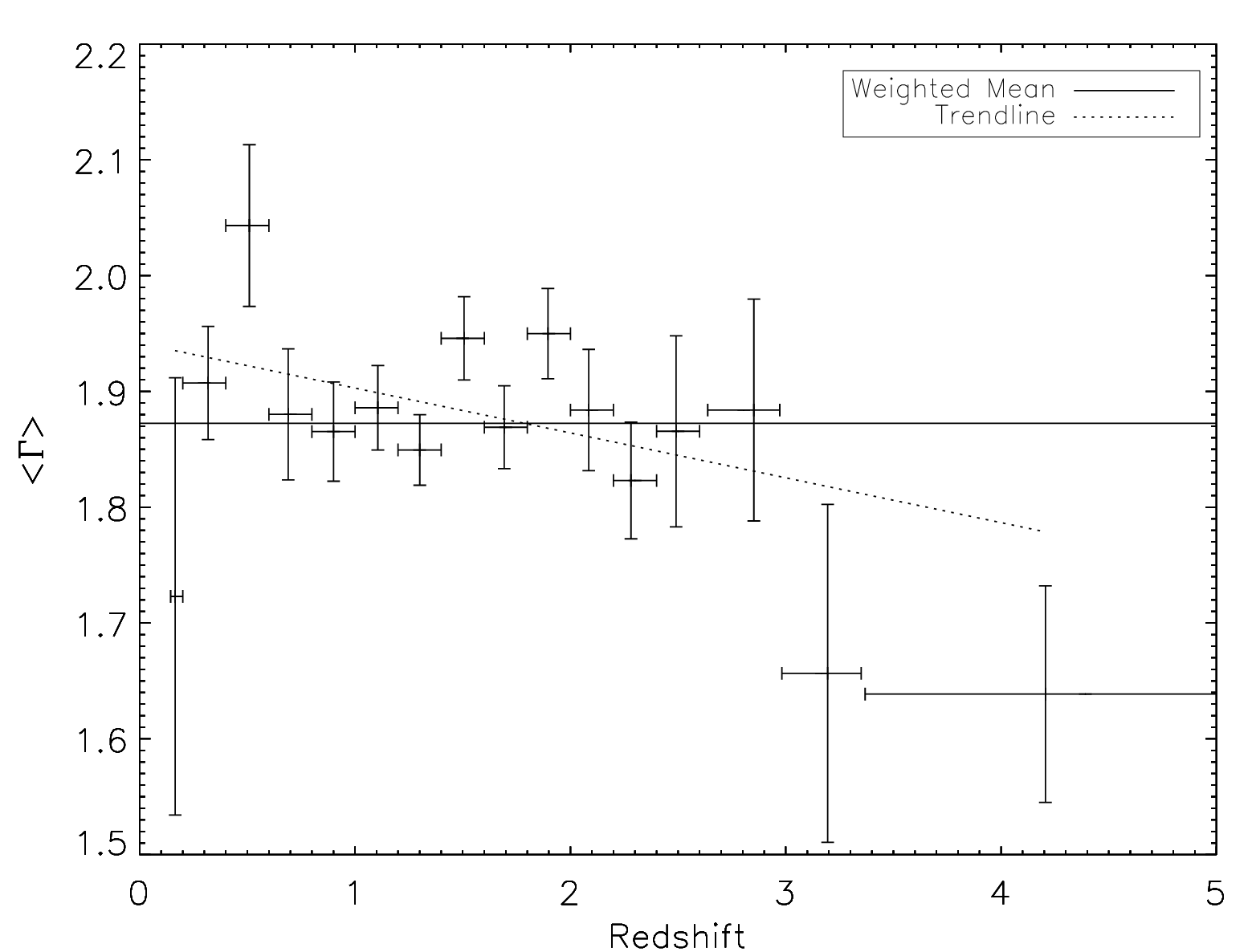}
  \includegraphics[width=0.5\textwidth]{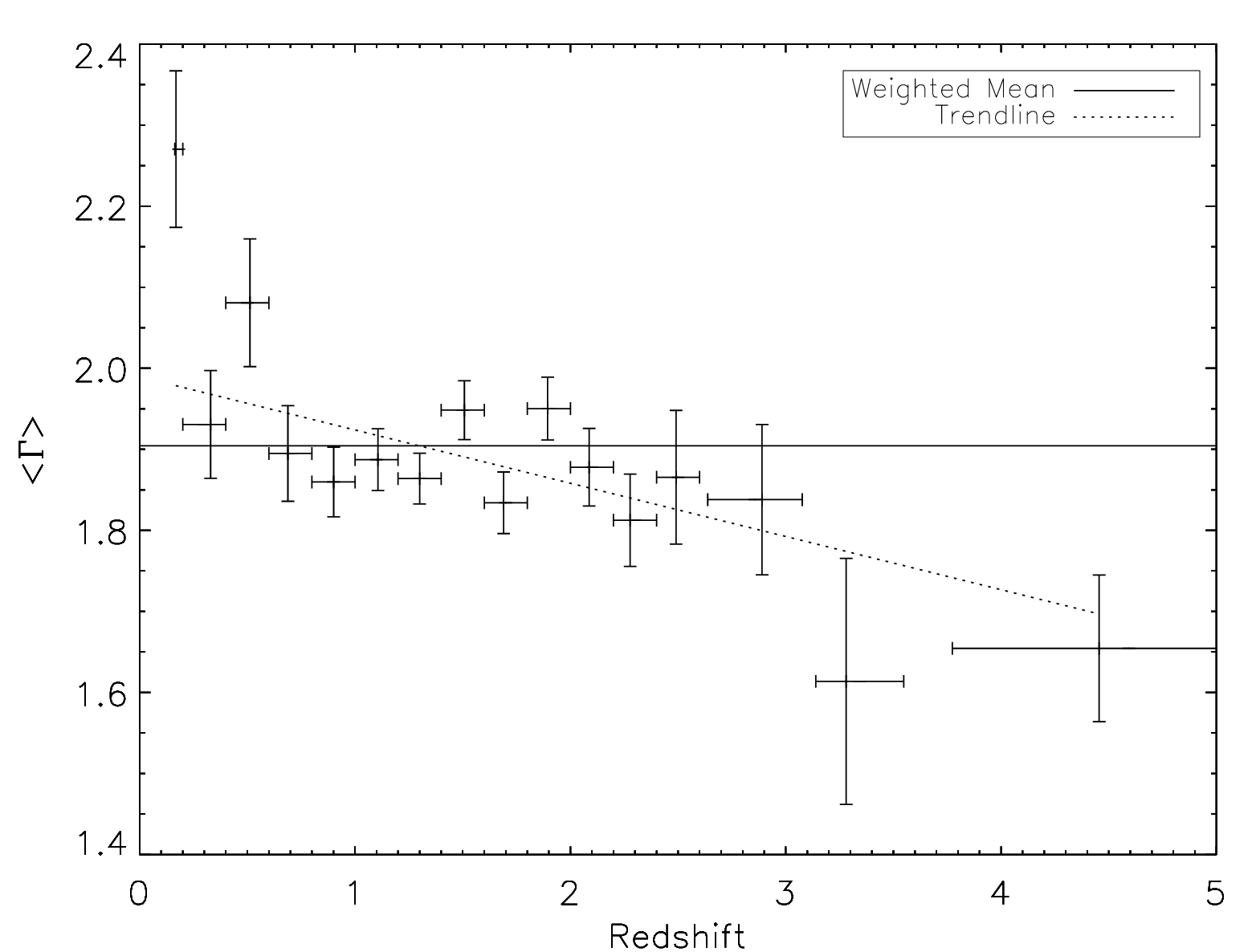}
  \caption{These plots show the dependence of weighted \gmm values on redshift.  The sources are binned in redshift (d\textit{z}=0.2), up until $z=2.6$ and then after this, to ensure at least 8 sources per bin (or 6 for the po only plot).  The y error bars are standard errors on the mean, $\alpha=\sigma/\sqrt{N}$, where the standard deviation is based on deviations from the weighted mean.  The x errors indicate the width of the bin.  The solid horizontal lines indicate the weighted mean values of each sample and the dotted lines shows the linear fits to the data.  The top plot includes data from 734 sources with good fits and a linear fit of $\left<\Gamma\right> = (-0.04 \pm 0.02)z + (1.94 \pm 0.03)$.  The bottom plot includes data from 650 sources with good fits using a simple power law model and a linear fit of $\left<\Gamma\right> = (-0.07 \pm 0.02)z + (1.99 \pm 0.03)$.}
  \label{fig:gmm_z_bin}
\end{figure}

Figure~\ref{fig:gmm_z_bin} shows binned versions of Figure~\ref{fig:gmm_z_scatter}.  The top plot includes sources fit with all models, whilst the bottom plot only considers sources best fit with the simple power law model.  This is considered because we find that a large percentage of sources in the redshift bins $z<0.6$ are those fit with the po+bb model, whereas the bins at higher redshift are mostly sources that are fit with the simple power law model.  Whilst the inclusion of the soft excess component should account for the spectral complexity present and give \gmm values that are consistent with those seen in the simple power law sources, this is not found for our sample.  The average \gmm value for the sources fit with the po+bb model is significantly lower than that of the entire sample as determined by a KS test. 

Narrow redshift bins (d\textit{z}=0.2) are used up to $z=2.6$.  Above this point, we bin so as to ensure at least 8 sources per bin due to decreased numbers of sources, (or 6 in the case of the po only plot).  A weighted mean value of \gmm is calculated for each bin according to the equation:

\begin{equation}
\label{eqn:weighted_mean}
\left<\Gamma\right> = \sum \rm{P}\sb{i} \times \Gamma\sb{i}   \,\,\,\,\, \rm{where} \,\,\,\,\, P\sb{i} = \frac{1 / \sigma\sb{i}\sp{2}}{\sum (1 / \sigma\sb{i}\sp{2})}
\end{equation}

which takes into account the individual errors ($\sigma\sb{i}$) on the \gmm estimates ($\Gamma\sb{i}$).  The errors on \wgmm are calculated as standard errors on the mean, $\alpha=\sigma/\sqrt{N}$, where:

\begin{equation}
\label{eqn:weighted_mean_error}
\sigma = \sqrt{\frac{1}{N-1} \sum (\Gamma\sb{i}-\left<\Gamma\right>)\sp{2}}
\end{equation}

\begin{table*}
\begin{minipage}{170mm}
\centering
  \caption{Listed are the best-fitting linear correlations between \gmm and $z$ or \lumx for the binned data, found by \chisq minimization.  The errors on the values are $1\sigma$.  The null hypothesis probabilites (p-values) for both the linear models, and constant \wgmm models are listed in column 4, where p$>1\%$ is taken to be an acceptable fit.}
  \label{table:gamma_corr}
     \centering
     \begin{tabular}{llll}
     \hline 
     Correlated Properties     & Sources Included     & Model                                                                     &  p-value      \\
     \hline
     \gmm and $z$              & All                  & $\left<\Gamma\right> = (-0.04 \pm 0.02)z + (1.94 \pm 0.03)$               &  14.2\%       \\
                               &                      & $\left<\Gamma\right> = 1.87 $                                             &  4\%          \\
                               & po only              & $\left<\Gamma\right> = (-0.07 \pm 0.02)z + (1.99 \pm 0.03)$               &  12\%         \\
                               &                      & $\left<\Gamma\right> = 1.90 $                                             &  0.012\%      \\
     \hline
     \gmm and \lumx            & All                  & $\left<\Gamma\right> = (-0.06 \pm 0.02)\rm{log} L\sb{X} + (4.6 \pm 1.1)$  &  38.6\%       \\
                               &                      & $\left<\Gamma\right> = 1.87 $                                             &  4.5\%        \\
                               & po only              & $\left<\Gamma\right> = (-0.15 \pm 0.03)\rm{log} L\sb{X} + (8.5 \pm 1.2)$  &  45.4\%       \\
                               &                      & $\left<\Gamma\right> = 1.90 $                                             &  $9\times 10\sp{-5}$\% \\
     \hline
  \end{tabular}
\end{minipage}
\end{table*}

We fit the data with 2 models; a linear trendline where a gradient inconsistent with zero would indicate \gmm values evolving with $z$, and a model where \wgmm was fixed at the value of the weighted mean for the sample to represent a non-evolving $\Gamma$.  The models and corresponding null hypothesis probabilites (p-values) can be found in Table~\ref{table:gamma_corr}. 

For the sources fit with all types of spectral model, a non-evolving \gmm gives a good fit to the data and the linear trend gives a gradient consistent with zero.  This suggests no significant trend for decreasing \gmm with increasing $z$.  However, when only the sources fit with a simple power law model are considered, the non-evolving \gmm model does not provide a good fit to the data and the linear trend gives a gradient inconsistent with zero.  This suggests that for this sub-set of sources, a significant, but slight, trend of \gmm with $z$ is present.

We note here that our estimate of \wgmm for the entire sample is lower than the arithmetic value determined without considering the individual errors on \gmm measurements.  This does not seem to be due to any particular source with vastly underestimated errors, but a general effect created because the slopes of flatter sources are easier to constrain and therefore have lower associated errors than the steeper sources.  If we remove the top 2.5\% of sources with weights above 0.01, we find the weighted mean value steepens to 1.93.

The lowest redshift bin of Figure~\ref{fig:gmm_z_bin} has a large error in \wgmm due to a low number of sources with a large scatter of \gmm values, which also causes the \wgmm value to change considerably between the 2 plots.  Similarly the higher redshift bins also include low numbers of sources and are likely unrepresentative of the entire sample as the percentage of sources that were targets of the \xmm observation is higher (for $z>2.6:25\pm11$\%, for $z<2.6:7\pm1$\%).  Therefore we investigate whether the lowest bin and top bins strongly influence the results and find that the overall conclusions remain the same, i.e. there is no dependence of \gmm on \textit{z} when all sources are considered, but a marginal trend is apparent when just those sources best fit with a simple power law are considered.

Since it is suggested that RLQ tend to have flatter spectra than RQQ, it is possible that an increased fraction of RL sources in high redshift bins may be a contributing factor to the flatter \gmm values observed. However, all the sources in our sample above $z=2.6$ are radio quiet using the definition of $R\sb{L}$ from Section~\ref{section:procedure}.

We note that when lower count sources are removed, the results for the binned data remain similar.

As our sample is flux limited, we are biased towards higher luminosity objects in higher redshift bins.  We therefore consider in Section~\ref{section:gmmlx} whether we are observing flatter power law slopes in higher luminosity objects, rather than flatter power law slopes in higher redshift objects. 

As we find a trend of \gmm with \textit{z}, we investigate whether this could be due to the presence of a reflection component which may be detected in higher redshift sources where the EPIC bandpass probes up to higher rest frame energies.  If this is not modelled separately, and just a simple power law model is used for the entire spectrum, its presence will make the slope of the power law appear flatter.  This could explain the slight decrease in the power law slopes seen for sources at higher redshift.  To test this hypothesis we fit simple power law models to spectra over the rest frame energy range 2.0-6.0 keV, with the upper limit chosen so as to exclude any contribution from such a reflection component and give a cleaner representation of the power law.  The lower limit of 2 keV is chosen to exclude any absorption component.  

\begin{figure}
  \centering
  \includegraphics[width=0.5\textwidth]{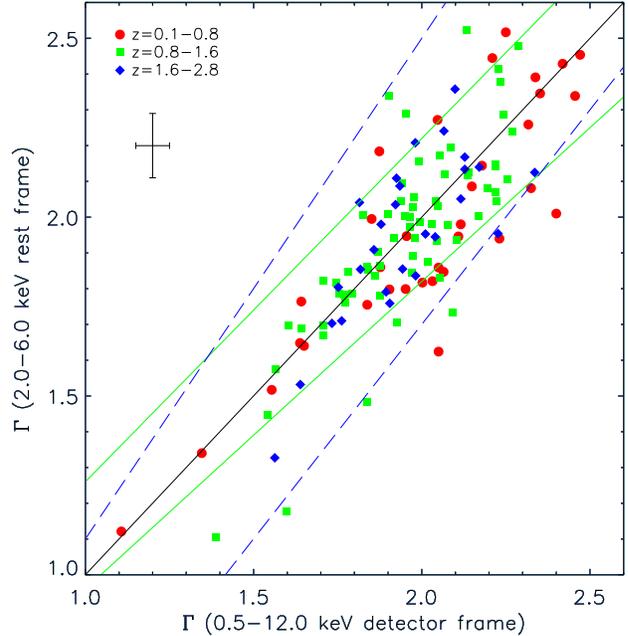}
  \caption{This figure plots the power law slope value as determined from a spectral fit over the entire 0.5-12.0 keV detector energy range against the value determined from a spectral fit over the reduced 2.0-6.0 keV rest frame energy range for 128 sources as detailed in the text.  They are divided into 3 broad redshift bins, red circles: $z=0.1-0.8$, green squares: $z=0.8-1.6$, blue diamonds: $z=1.6-2.8$.  An indication of the typical 68\% error is shown offset from the real data points.  The line where \gmmpart = \gmmfull is marked on the figure for reference and does not represent a particular linear fit.  The solid green/dashed blue lines represent 68\% error bounds on the best-fitting trendline for the $z=0.8-1.6$/$z=1.6-2.8$ sources respectively.  It can be seen that all sources lie within these boundaries, suggesting the sub-samples are consistent with each other.}
  \label{fig:wgmm}
\end{figure}

Figure~\ref{fig:wgmm} plots the power law slope value measured over the reduced range, 2.0-6.0 keV in the rest frame ($\Gamma\sb{\rm{part}}$) against the slope measured over the full 0.5-12.0 keV observed frame ($\Gamma\sb{\rm{full}}$), for a reduced sample of 128 sources previously best-fit with a simple power law, and including $>250$ counts over the reduced range in each camera that is in use.  They are divided into 3 broad redshift bins and the typical 68\% error size is indicated to the left of the figure. If we are removing an unmodelled reflection component when considering the spectral fits over the reduced range, we would expect \gmmpart to be steeper than \gmmfull and hence the sources should lie in the top left corner, above the \gmmpart = \gmmfull line.  We calculate the \chisq statistic between the points and the null hypothesis line of \gmmpart = $\Gamma\sb{\rm{full}}$.  We find all 3 redshift bins to be consistent with this line indicating that there is no significant difference between \gmmfull and $\Gamma\sb{\rm{part}}$.  This suggests no strong reflection component is present in these sources and therefore cannot be used to explain the flattening of \gmm with increasing $z$. 

   \subsection{Luminosity Dependence of the Power Law Slope}
   \label{section:gmmlx}

\begin{figure}
  \centering
  \includegraphics[width=0.48\textwidth]{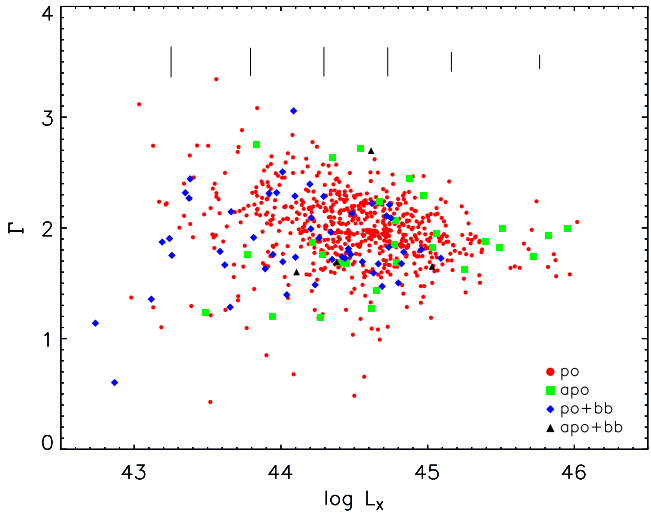}
  \caption{This plot includes the 734 sources with H0$>1\%$.  The different colours/symbols correspond to the best fitting model; red circles are po, green squares are apo, blue diamonds are po+bb and black triangles are apo+bb.  The error bars shown at the top of the plot correspond to the average of each of the $1\sigma$ errors on the individual sources.}
  \label{fig:gmm_lx_scatter}
\end{figure}

In this Section we consider whether our data shows any evidence for an intrinsic variation of the spectral index, $\Gamma$, as a function of hard (2-10 keV) X-ray luminosity which is plotted in Figure~\ref{fig:gmm_lx_scatter}.  We use the hard band since this energy range is less affected by intrinsic absorption.

We calculate Spearman's rank correlation coefficient, $\rho$, for both the entire population of 734 sources, and the subset of 650 sources whose best fit is a simple power law model (po).  Considering the entire population we find a negative $\sim8\sigma$ correlation of \gmm with log $L\sb{X}$ ($\rho$ = -0.24, prob = $4 \times 10\sp{-11}$).  Considering only the sources fit by po, we find a tighter negative $\sim 8\sigma$ correlation ($\rho$ = -0.29, prob = $2 \times 10\sp{-14}$).  These correlations imply that it is the sources with higher hard X-ray luminosities that tend to have flatter \gmm values.   

As previously discussed, the higher redshift objects in the sample may be present due to pre-selection.  Since these sources are also likely to be the sources with the highest luminosities, we test whether their exclusion from the sample has any effect on the correlation coefficients.  We find that when only considering sources with $z<2.6$, the correlation of \gmm with \lumx remains present at $8\sigma$.

\begin{figure}
  \centering
  \includegraphics[width=0.5\textwidth]{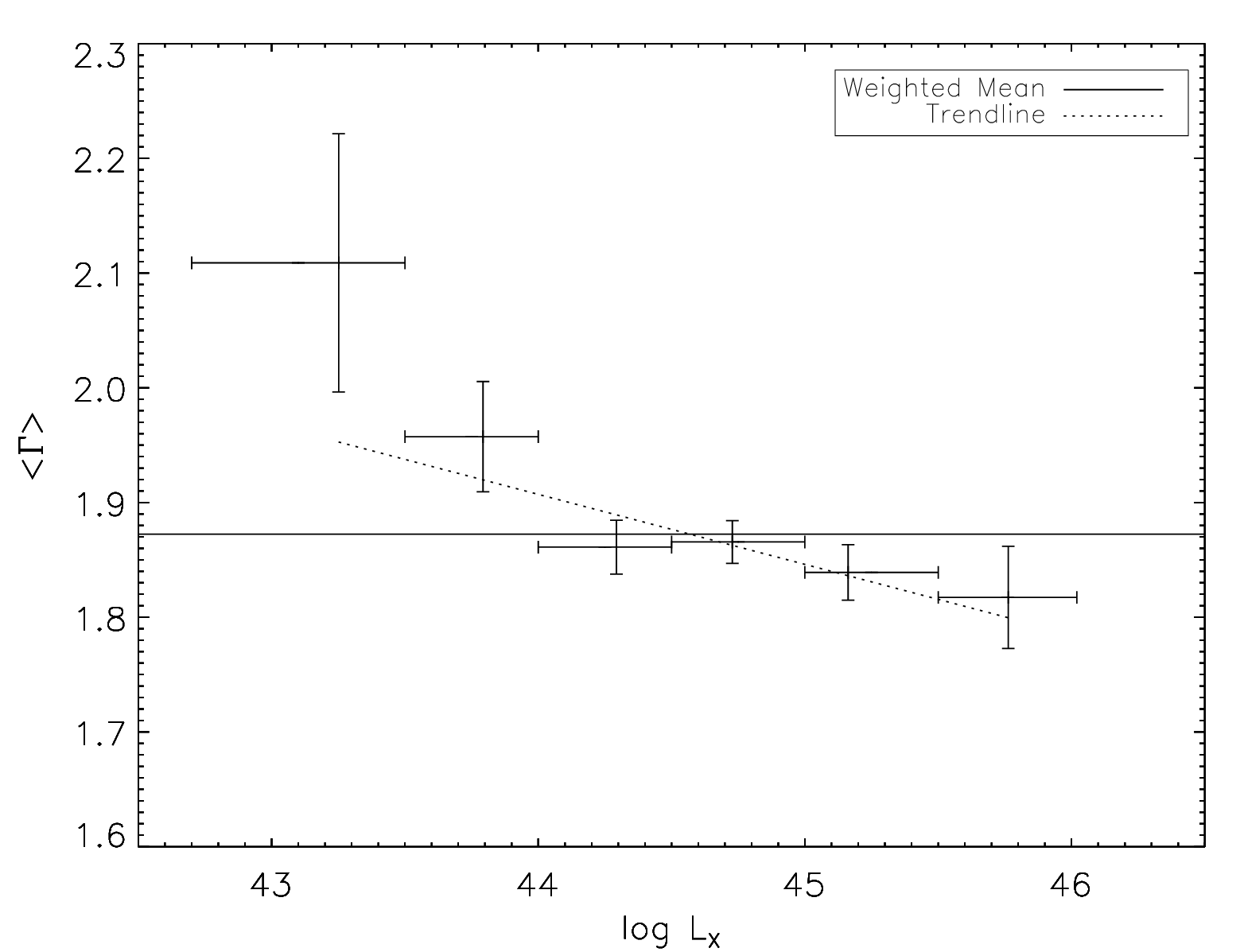}
  \includegraphics[width=0.5\textwidth]{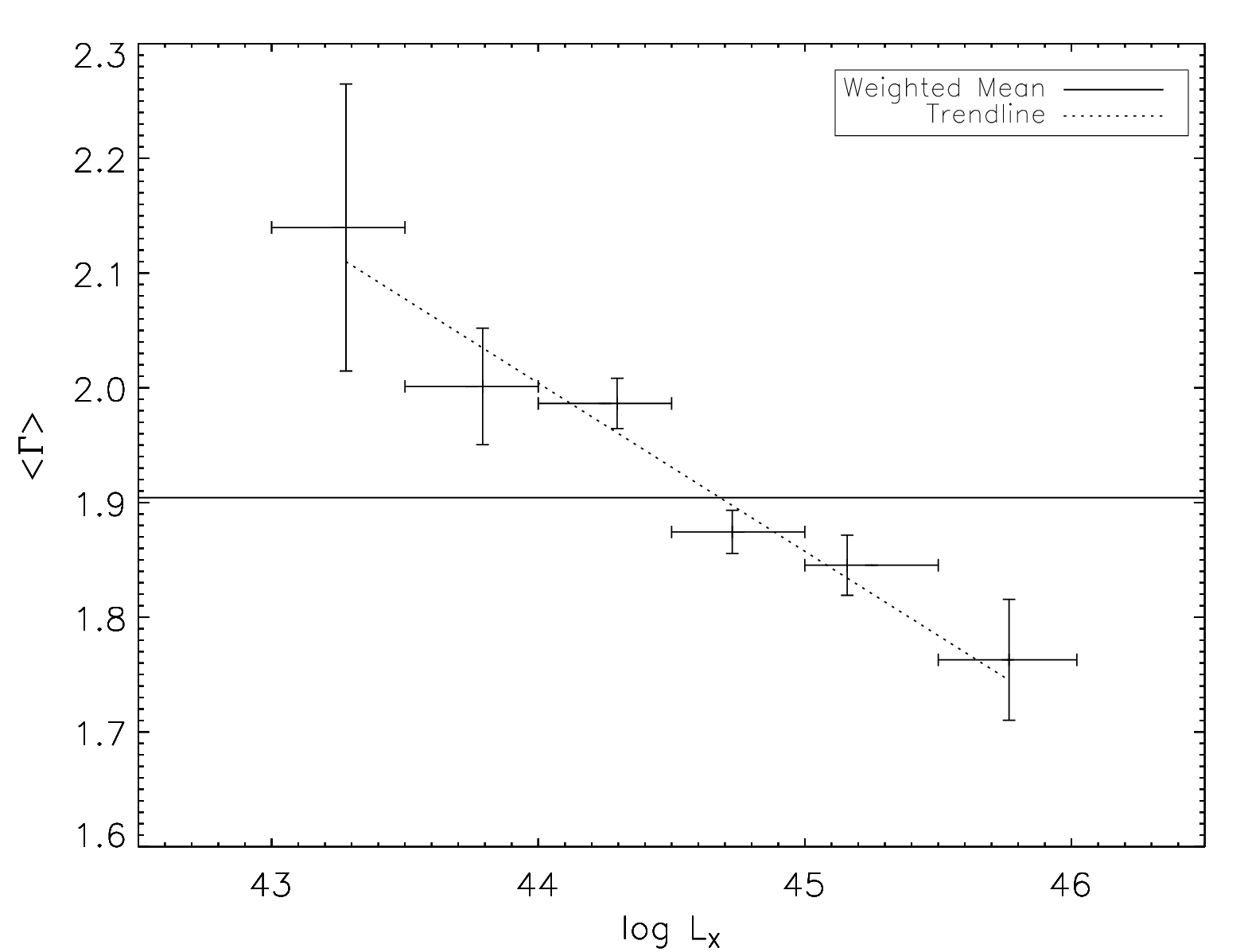}
  \caption{These plots show the dependence of weighted \gmm values on hard (2-10 keV) luminosity.  The sources are binned in approximately log L$\sb{X}$ = 0.5 bins, although the lower and upper bins do include sources slightly below 43.0 and above 46.0 respectively.  The solid horizontal lines indicate the weighted mean values of each sample and the dotted lines show the linear fit to the data. The top plot includes data from 734 sources with good fits using models with freely varying \gmm values and a linear fit of $\left<\Gamma\right> = (-0.06 \pm 0.02)\rm{log}L\sb{X} + (4.6 \pm 1.1)$.  The bottom plot includes data from 650 sources with good fits using the simple power law model and a linear fit of $\left<\Gamma\right> = (-0.15 \pm 0.03)\rm{log}L\sb{X} + (8.5 \pm 1.2)$.  The y error bars are standard errors on the mean and the x errors indicate the width of the bin. }
  \label{fig:gmm_lx_bin}
\end{figure}

Figure~\ref{fig:gmm_lx_bin} shows a binned version of Figure~\ref{fig:gmm_lx_scatter}, with the top plot including sources fit with all models, and the bottom plot including just those fit with the simple power law model.  Each bin size is approximately log L$\sb{X}$ = 0.5, and includes at least 18 sources.  A weighted mean value of \gmm is calculated for each bin using Equation~\ref{eqn:weighted_mean} as before.

As in the previous section, we calculate linear trends to the data using \chisq minimization and also consider whether the data are consistent with a \wgmm that is constant with $L\sb{X}$.  These trends along with the null hypothesis probabilities for each model can be found in Table~\ref{table:gamma_corr}.

Considering sources best fit with all types of spectral model, the linear trend gives a gradient only consistent with zero at $3\sigma$.  However, a constant \gmm also gives an acceptable fit to the data meaning that a non-evolving \gmm model cannot be ruled out.  When only the sources fit with a simple power law model are considered, the non-evolving \gmm model does not provide a good fit to the data and the linear trend gives a gradient inconsistent with zero.  This suggests that for this sub-set of sources, a significant trend of \gmm with \lumx is present.

We also investigate whether the trends we observe are still present in the data when we restrict the sample to sources above certain count levels.  For the un-binned data, a significant correlation is still present between \lumx and \gmm even when all the sources with $<1000$ counts have been removed, although the significance level does fall.  The binned data shows similar results for all count levels i.e. a trend that is only consistent with zero at $3\sigma$ (all models) and $5\sigma$ (po models).

We test whether the correlation we observe is still present when the RLQs are removed from the sample, since it has been suggested that they have flatter spectra than RQQ.  A significant trend between \gmm and \lumx is still seen (5$\sigma$).

Since we find the stronger of the two correlations is between \gmm and $L\sb{X}$, we suggest that it is luminosity that is the dominant variable, rather than redshift.  We expect such a correlation from a consideration of the spectral shapes and how the luminosity is calculated.  For example, a source with a steep spectral slope will have less emission over the hard X-ray range, compared with a source with a flat spectral slope.


\subsection{Radio Properties}
\label{section:radio}

The radio loudness values ($R\sb{L}=F\sb{R}/F\sb{O}$) of our sources were determined using radio fluxes from FIRST \citep{FIRST} and the optical fluxes from SDSS \citep{DR5} as previously described in Section~\ref{section:procedure}.  Using the classification of \citet{kellermann89} where $R\sb{L} > 10$ indicates the source is radio loud, 552 sources were confirmed as radio quiet and 75 as radio loud.  The properties of these sub-samples are investigated in this section. 

\begin{figure}
  \centering
   \includegraphics[width=0.38\textwidth]{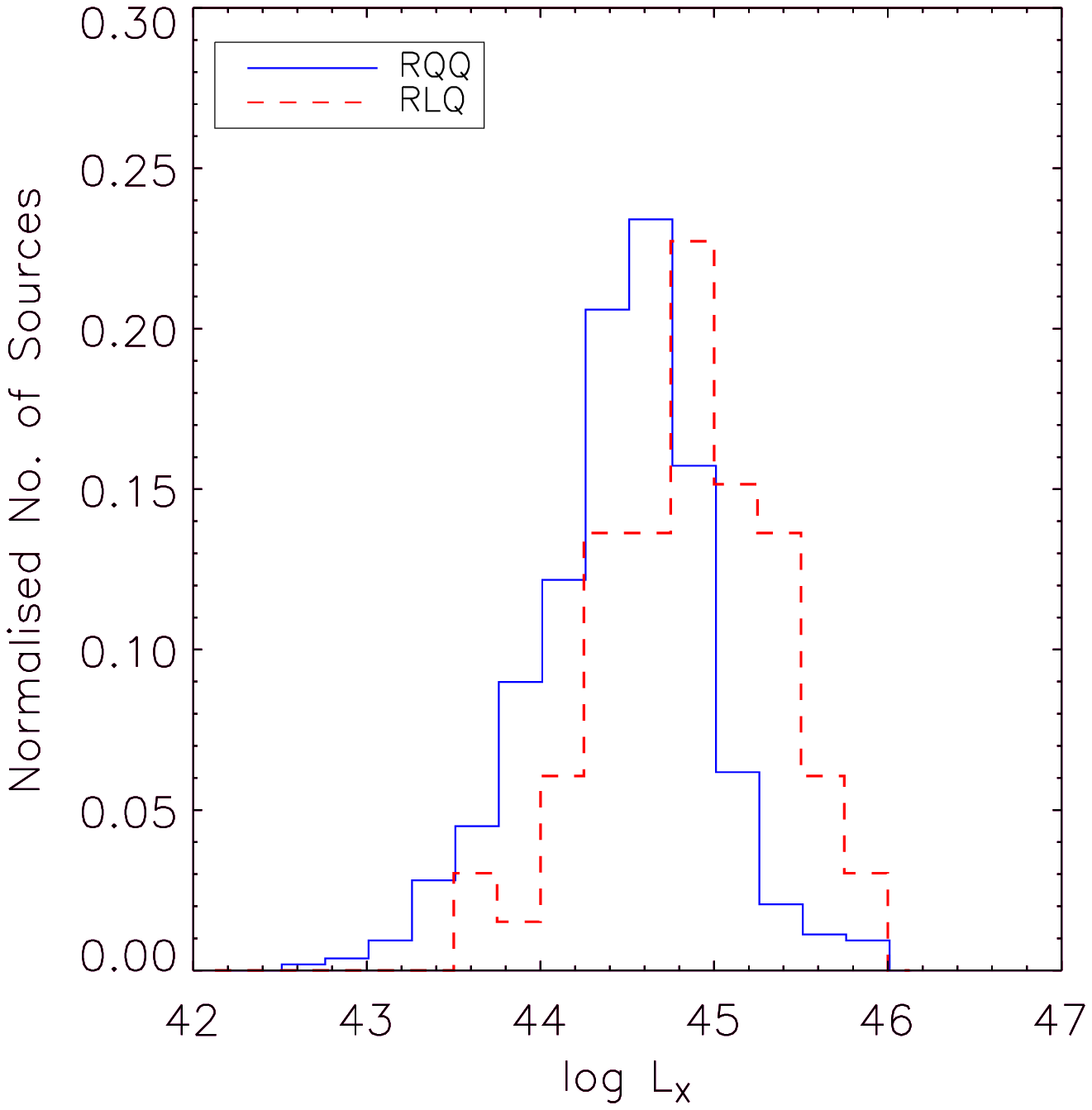} \\
   \vspace{3mm}
   \includegraphics[width=0.38\textwidth]{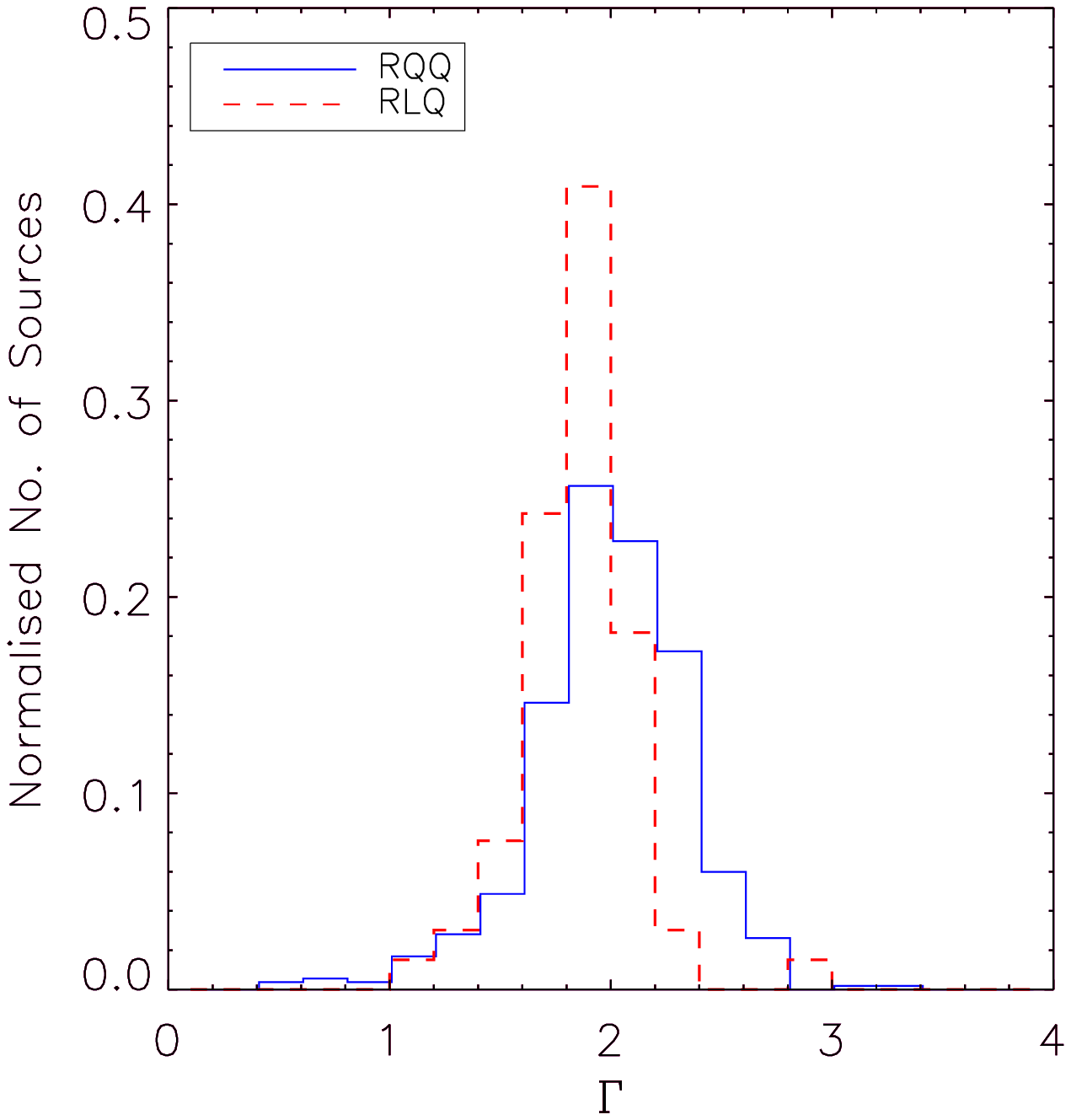} \\
   \vspace{3mm}
   \includegraphics[width=0.38\textwidth]{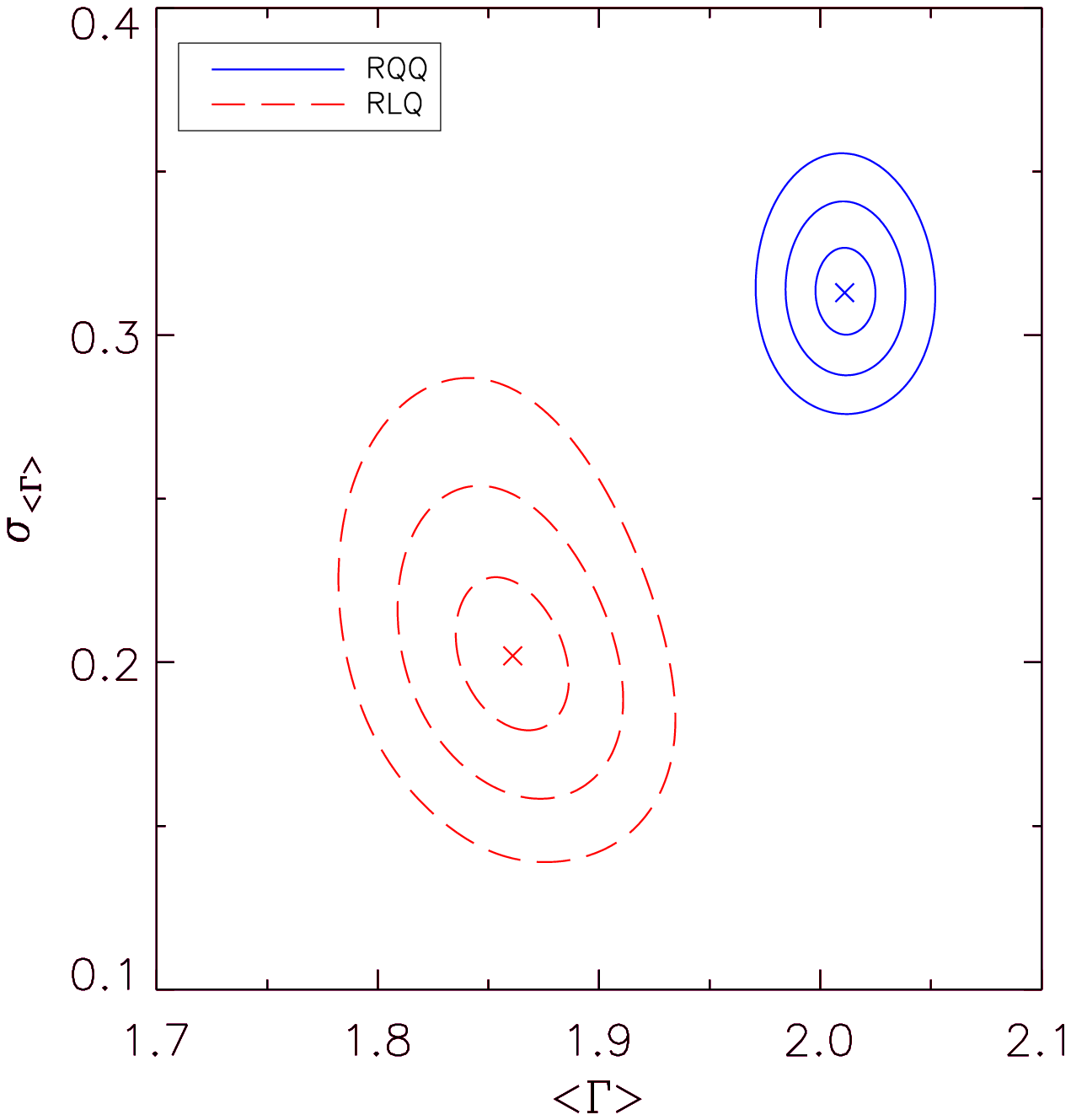}
  \caption{These plots compare the \lumx (top) and \gmm (middle) distributions of the confirmed radio quiet (solid blue) and radio loud (dashed red) sources.  The bottom figure shows the best-fitting \wgmm and \sigwgmm values (crosses) and 1$\sigma$, 2$\sigma$ and 3$\sigma$ confidence contours from the Gaussian fitting of the \gmm distributions for the RQQ and RLQ sub-samples.}
  \label{fig:radio_distributions}
\end{figure}

The distributions of absolute \textit{i} band magnitudes for the RQQ and RLQ sub-samples are found to not be significantly different (KS significance = 0.39).  In addition the redshift distributions are not significantly different (KS significance = 0.87).  The X-ray luminosity distributions for RQQ and RLQ presented in Figure~\ref{fig:radio_distributions} (top) are found to be significantly different (KS significance = $8\times10\sp{-9}$), with RLQ having higher 2-10 keV X-ray luminosities.  Since the redshift distributions are the same, this is likely due to a physical difference in the sources rather than an observational bias from the sample being flux limited.  This suggests that an additional mechanism of producing X-rays is present, which increases the total X-ray emission.  This idea has been previously suggested by \citet{reeves00}.  The additional component is perhaps related to the jet and may be Synchtrotron Self Compton (SSC) scattering where the radio photons produced as synchrotron emission from electrons spiralling in a magnetic field, undergo inverse Compton scattering off the same electrons, up to X-ray energies \citep{SSC}.

It is thought that the synchrotron emission produced in RLQ will contaminate the power law produced by inverse Compton scattering, making the X-ray spectrum appear flatter.  We therefore explore the differences in the \gmm distribution of the sub-samples of RLQ and RQQ, as shown in Figure~\ref{fig:radio_distributions} (middle) and a KS test finds that they are significantly different (KS significance$=3\times10\sp{-5}$).  When modelled with a single Gaussian, the RQQ distribution has a mean value of \gmm $= 2.01 \pm 0.01$, whereas the RLQ distribution has a lower mean value of \gmm$ = 1.86 \pm 0.02$, however we note that the 2 populations cover a similar range of \gmm values.  Flatter \gmm values in RLQ are noted in the literature, although the actual values found here are steeper than those reported (e.g. \citealt{reeves00}).

We also note that the dispersion of \gmm values is much smaller for the RLQ than the RQQ.  We find the intrinsic dispersions to be \sigwgmm = $0.31 \pm 0.01$ for the RQQ, which is similar to that for the entire original sample and \sigwgmm = $0.20 \pm 0.02$ for the RLQ, which is considerably lower.  The typical 68\% error on \gmm for the RLQ is 0.08, significantly lower than the intrinsic dispersion.  Figure~\ref{fig:radio_distributions} (bottom) shows the best-fitting \wgmm and \sigwgmm values and 1$\sigma$, 2$\sigma$ and 3$\sigma$ confidence contours from the Gaussian fitting of the \gmm distributions.  This clearly shows that the \gmm distribution for RLQ has a lower dispersion than that for RQQ and the average \wgmm value is also lower.

\begin{figure}
  \centering
   \includegraphics[width=0.48\textwidth]{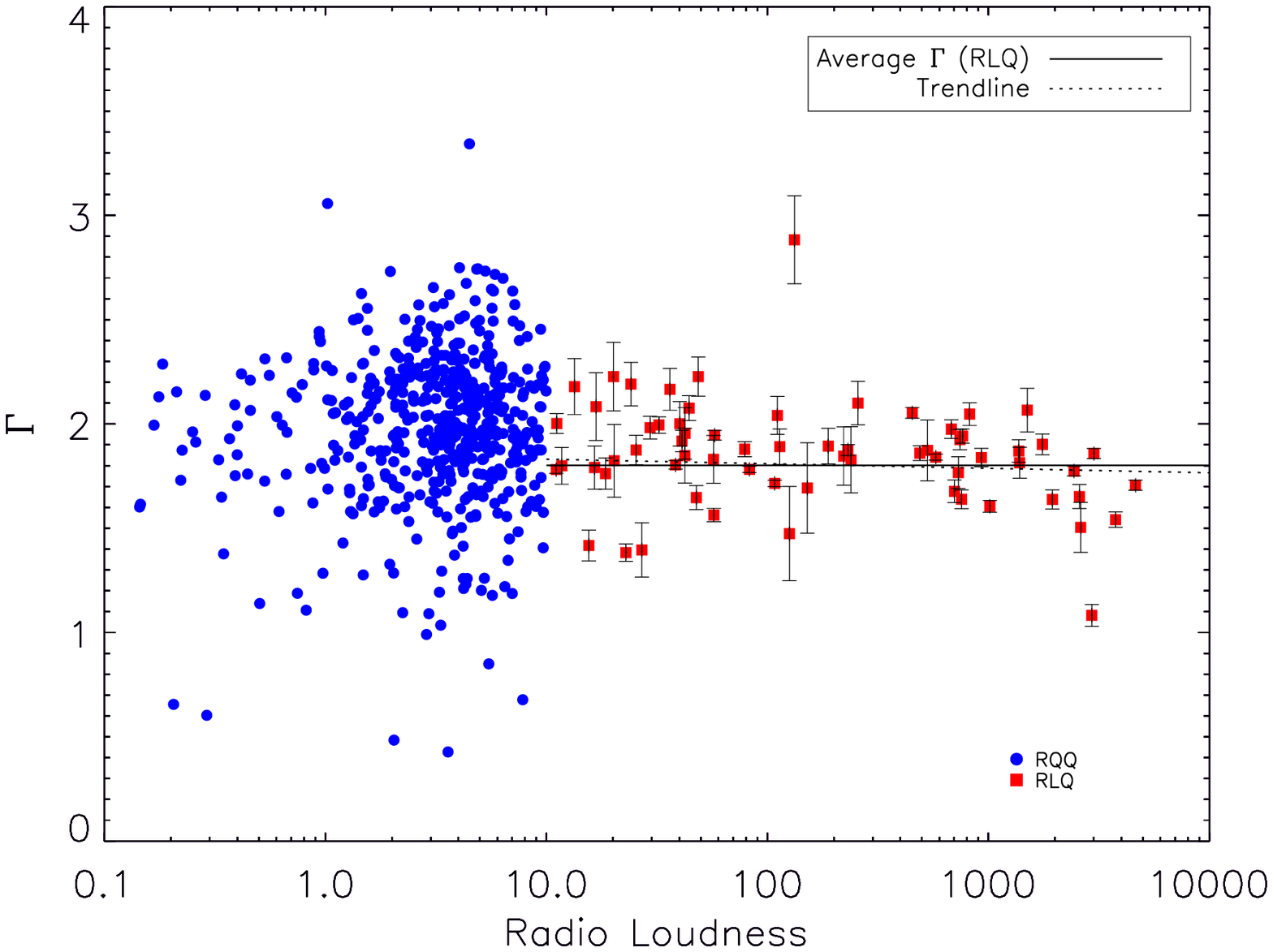} \\
   \vspace{3mm}
   \includegraphics[width=0.48\textwidth]{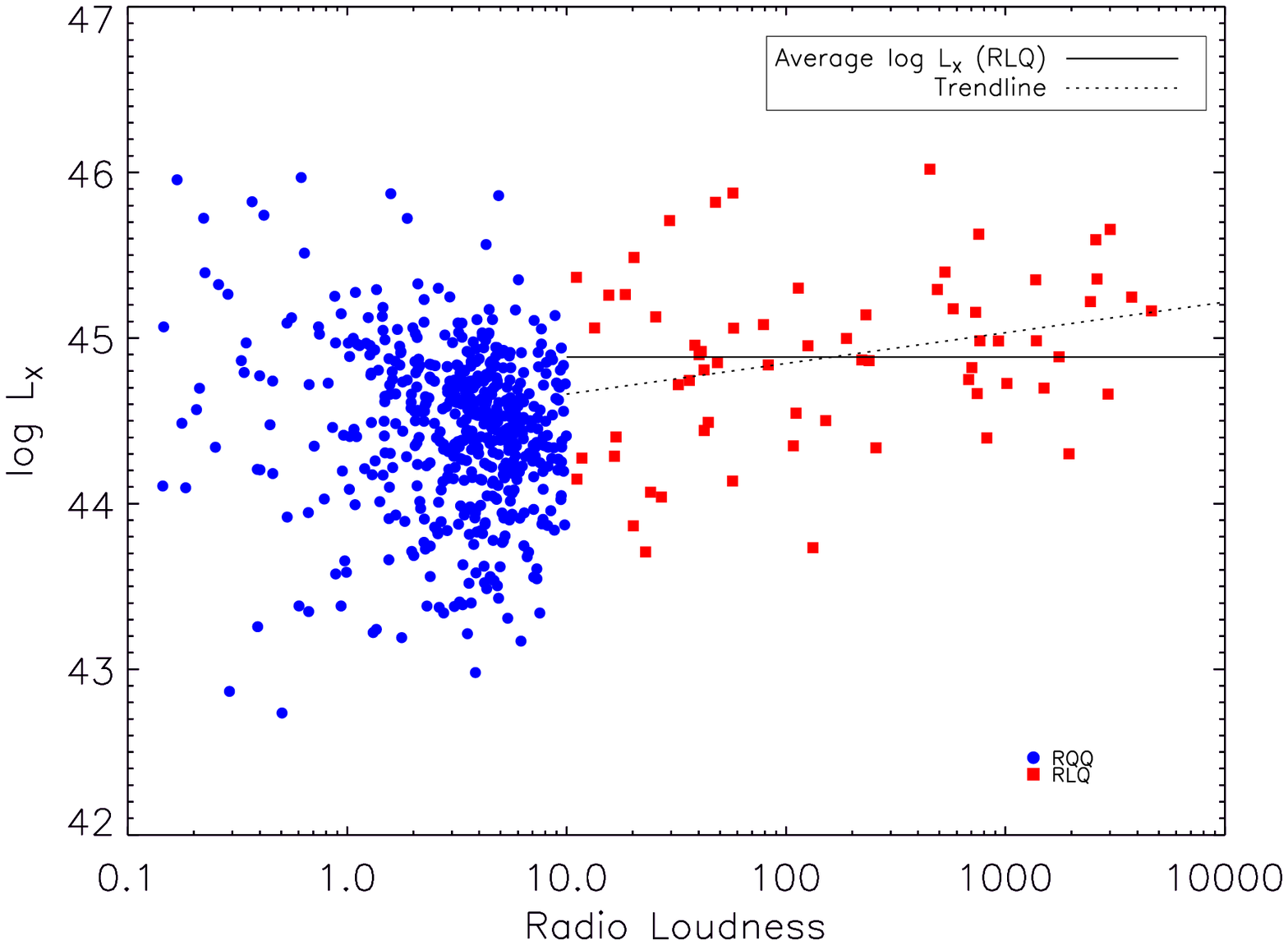} 
  \caption{This plots the radio loudness, $R\sb{L}$, as a function of the best-fitting \gmm values (top) and log \lumx (bottom).  The RLQ are shown as red squares and the errors on \gmm correspond to the 68\% confidence level.  The RQQ are shown as blue circles, for which the errors on \gmm have been ommited for reasons of clarity.  The solid lines show the average \gmm (1.80) and log \lumx (44.88) values for the RLQ sample, which represent the non-evolving models.  The dashed lines show the best-fitting linear trends to the RLQ: \gmm = $(-0.022 \pm 0.006)\textrm{log} R\sb{L} + (1.85 \pm 0.02)$ and log($L\sb{X}$) = $(0.17 \pm 0.002)\textrm{log} R\sb{L} + (44.5 \pm 0.003)$.}
  \label{fig:radioloudness}
\end{figure}

Figure~\ref{fig:radioloudness} (top) plots the radio loudness parameter $R\sb{L}$ against $\Gamma$.  The solid line shows the average \gmm value for the RLQ sample whilst the dashed line shows the best-fitting linear trend to the RLQ.  We expect to see flatter \gmm values for increasing levels of radio loudness as suggested from Figure~\ref{fig:radio_distributions} (middle) and the idea that the power law will become contaminated by emission from the radio jet.  We find the data is inconsistent with a constant \gmm fit (p=0\%), but an F-test shows that the linear fit is not statistically required.  This is also the case when the combined RQ and RL populations are considered.  Since both the constant and linear fits give poor fits to the data due to the large spread in values, we also consider the data when it is binned by radio loudness.  In this case the data is well fit by the constant \gmm model (p=15.3\%) and an F-test does not require the linear model.  This is again also true for the combined RQ and RL populations and therefore we do not report a significant trend between $R\sb{L}$ and $\Gamma$.  

Figure~\ref{fig:radioloudness} (bottom) plots the radio loudness parameter $R\sb{L}$ against $L\sb{X}$.  The solid line shows the average $L\sb{X}$ value for the RLQ and the dashed line shows the best-fitting linear trend to the RLQ sample.  The result from Figure~\ref{fig:radio_distributions} (top) suggests that the more radio loud objects also tend to be more X-ray luminous.  The data is inconsistent with a constant luminosity model (p=0\%), but a linear fit is not statistically required.  We note that for the combined RQ and RL populations, the linear trend is required, but this model still does not provide a good fit to the data (p=0\%).  When we consider binned data, the constant log \lumx model is still a poor fit to the RLQ ($\textrm{p}=5 \times 10\sp{-11}$), but the linear model is now statistically required.  However, the gradient of the linear trend is $0.17 \pm 0.08$ which is only significant at $\sim 2\sigma$.  For the combined RQ and RL populations, the linear trend is not required for the binned data.

   \subsection{Intrinsic Cold Absorption}
   \label{section:apo}

Whilst the majority of our sources were well fit by a simple power law model, some sources were better fit with an absorbed power law which includes an additional intrinsic neutral absorption component at the redshift of the source (modelled by \texttt{zphabs} in \texttt{XSPEC}).  The requirement of this component was tested by the F-test at 99\% significance.

In the sample of 734 sources which have H0$>$1\%, 32 sources required an additional absorption component.  Setting the F-test confidence at the 99\% level, requires that we take into account the presence of 1\% of detections which could be spurious.  Accounting for these gives 3.4\% absorbed sources.  This represents only a lower limit to the number of absorbed sources since the detection of absorption at the required significance is more difficult in spectra with lower numbers of counts.  This value is lower than those in the literature, with many studies reporting absorbed fractions of $\sim 10\%$, albeit for lower F-test significances \citep{corral11, young09, mainieri07, mateos05b, mateos05a}. \citet{mateos10} uses the more stringent significance threshold of 99\% as we do here, but still find a greater absorbed fraction of $\sim 8\%$.  

\begin{figure}
  \centering
  \includegraphics[width=0.5\textwidth]{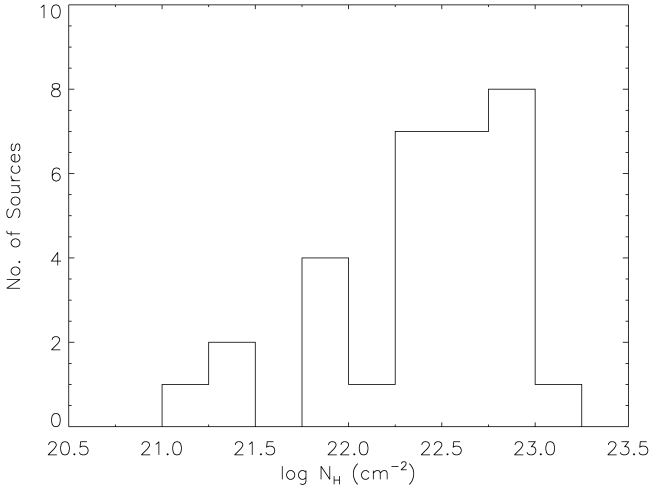}
  \caption{The distribution of intrinsic absorbing column densities for the 32 sources in the sample that require the additional absorption component.}
  \label{fig:nh_dist}
\end{figure}

Both the literature results and ours found here are at odds with the standard orientation based Unified Model \citep{antonucci93}, which does not predict sources that are optically classified as type 1 to require an absorbed power law model in X-rays.  The typical levels of absorption in our sources range from $10\sp{21}$cm$\sp{-2}$ to $10\sp{23}$cm$\sp{-2}$, far higher than is expected for an AGN defined as type 1.  Figure~\ref{fig:nh_dist} shows the distribution of $N\sb{H}\sp{intr}$ values for all the absorbed sources in our sample.  A visual inspection of the optical spectra of the absorbed sources hints that some may be of an intermediate type (e.g. Sy 1.5), and others may be BALQSOs, offering a possible explanation for their apparent high levels of absorption.  An analysis of the optical reddening properties of the sample will be discussed in Hutton et al. (in prep).

\begin{figure}
  \centering
  \includegraphics[width=0.5\textwidth]{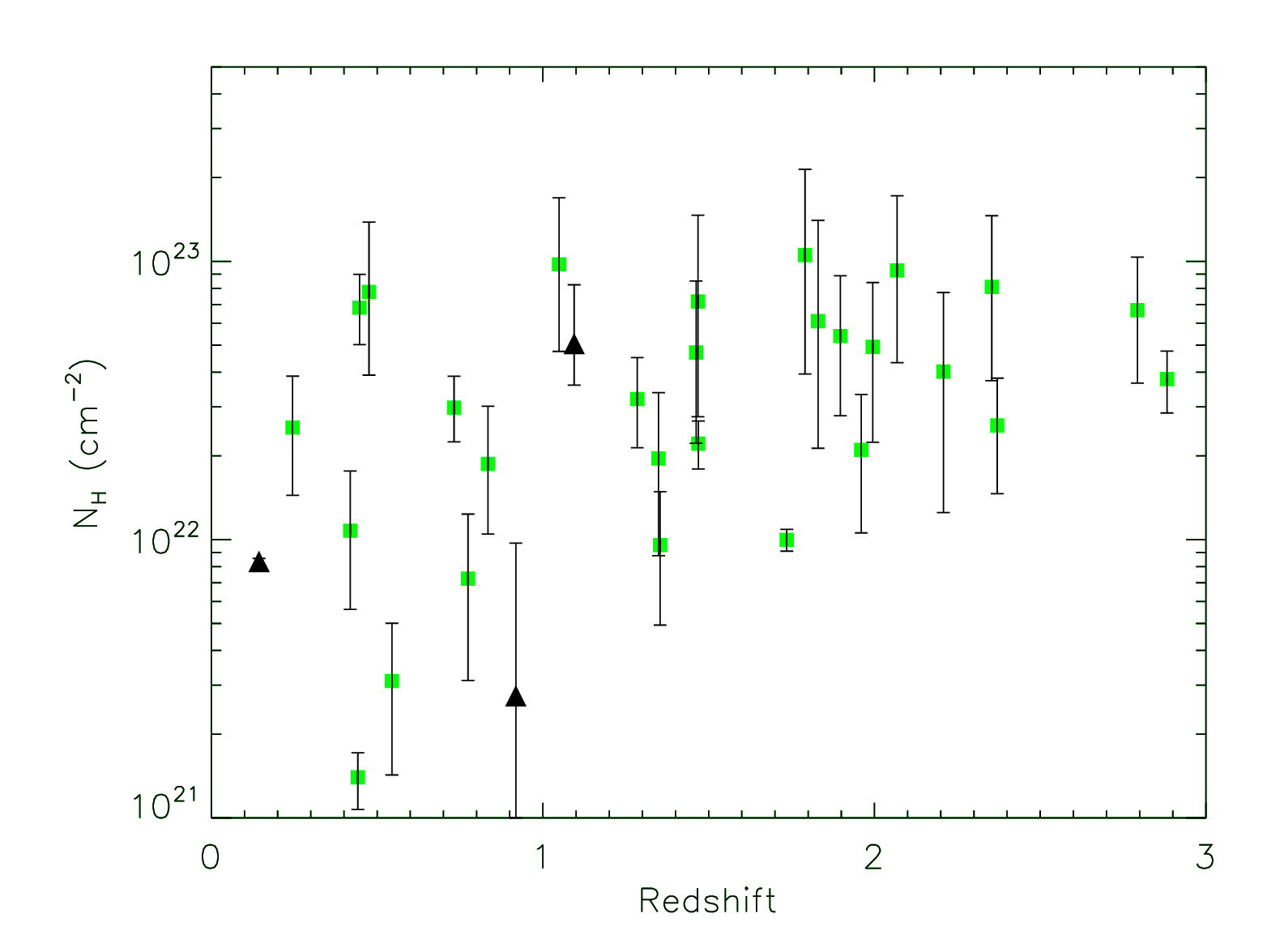}
  \includegraphics[width=0.5\textwidth]{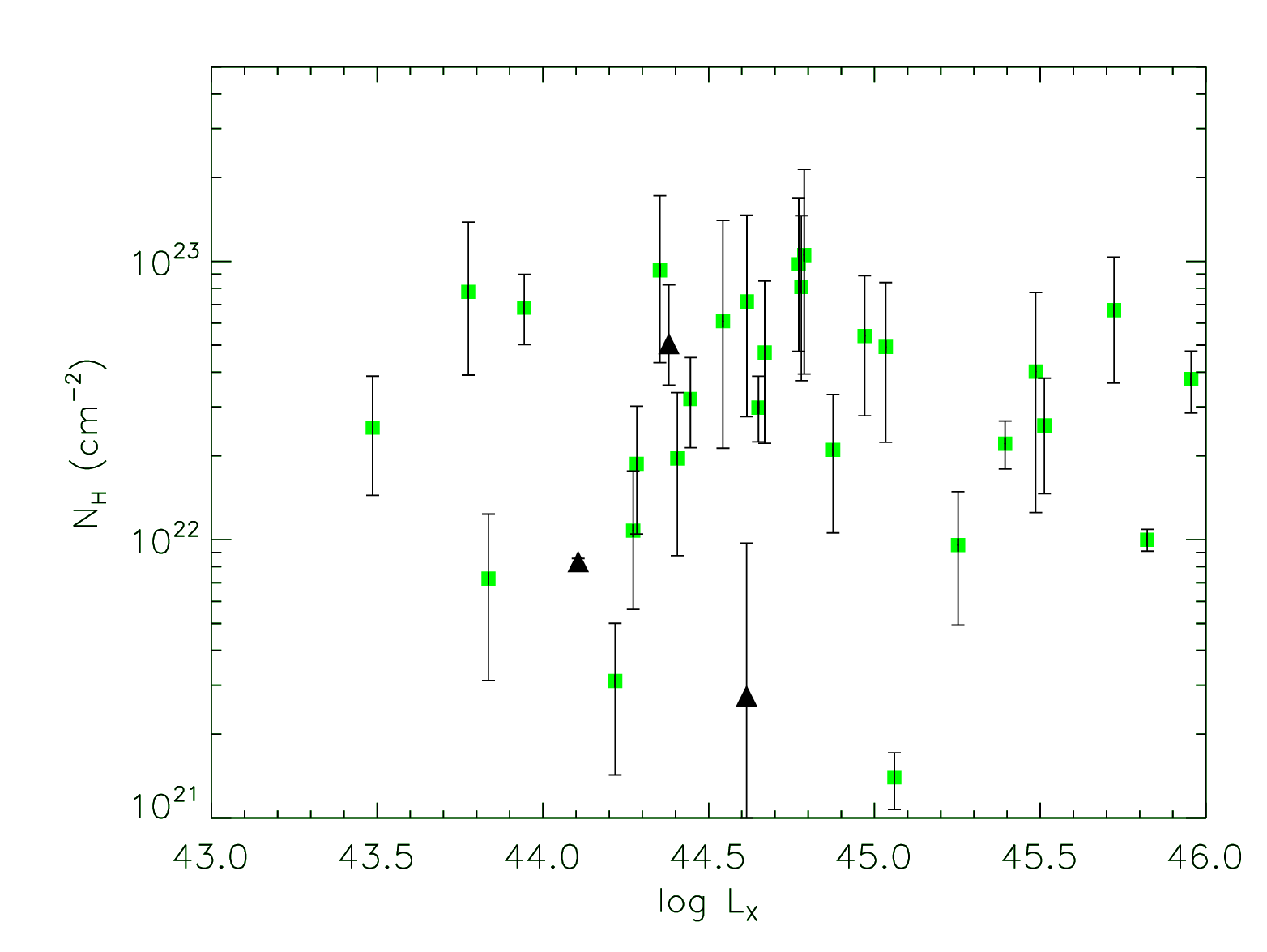}
  \caption{The relations between the intrinsic absorbing column density and redshift (top) and 2-10 keV X-ray luminosity (bottom). Green squares indicate the sources best fit with the apo model and black triangles indicate the apo+bb model.  The error bars correspond to the 68\% confidence range.}
  \label{fig:nh_trends}
\end{figure}

In Figure~\ref{fig:nh_trends} (top) we investigate the dependence of the intrinsic column densities of the absorbed sources, measured in the rest frame of the source, on redshift.  Although we see a slight trend for higher absorption levels in higher redshift sources, this can be explained as an observational bias, since sources with low amounts of absorption will not be detected at higher redshifts due to their absorption signature falling outside of the EPIC bandpass.  We do not detect absorption in any sources above $z=3$ where the lower energy limit of the \xmm bandpass corresponds to a source energy of $\sim 2$ keV.  At $z=2$, where this limit falls to $\sim 1.5$ keV, some absorbed sources become detectable, but only at levels high enough to create an absorption signature that falls into the observed bandpass.  Considering correlation coefficients, we find no significant trend of intrinsic column density with redshift (prob$=0.02$).

We also consider the relationship between $N\sb{H}\sp{intr}$ and hard X-ray luminosity in Figure~\ref{fig:nh_trends} (bottom).  We consider the X-ray luminosity in the hard band (2-10 keV) rather than the soft band (0.5-2.0 keV) as this is the energy range at which any intrinsic absorption present in the source would suppress the emission.  There is no correlation between $N\sb{H}\sp{intr}$ and \lumx which is confirmed by correlation coefficients (prob$=1.0$).

\begin{figure}
  \centering
  \includegraphics[width=0.5\textwidth]{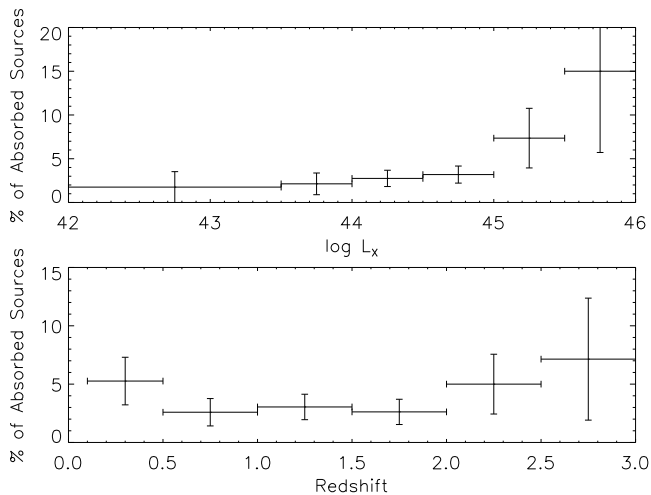}
  \caption{These figures show how the percentage of sources that require an intrinsic absorption component in their best-fitting model varies for different luminosity and redshift bins.  The x error bars indicate the width of the bins, and the y error bars come from Poissonian counting statistics.}
  \label{fig:abs_frac}
\end{figure}

We investigate how the fraction of absorbed sources changes with increasing luminosity and redshift.  This can be seen in Figure~\ref{fig:abs_frac} and the fraction is found to be constant across the different luminosity and redshift bins.  Previous studies (e.g. \citealt{hasinger08}) which include both `unabsorbed' type 1 and `absorbed' type 2 objects find the percentage of absorbed sources decreases with increasing luminosity, which supports the Receding Torus model of \citet{lawrence91}.  This notes that the dust sublimation radius increases as a function of luminosity and therefore the inner edge of the torus will be further away for higher luminosity sources, resulting in less obscuration.  Although our sample includes only type 1 objects, a small percentage of these have been shown to include intrinsic absorption, which might be expected to behave in a similar way to the absorption found in type 2 objects.  However, since we show the percentage of absorbed objects to remain constant with increasing luminosity, the receding torus model is not supported and the absorption present in our type 1 sources must be of a different nature.

   \subsection{Soft X-ray Excesses}
   \label{section:po+bb}

The sources were also fit to include a soft excess component, modelled with \texttt{zbbody} in \xspec (see Section~\ref{section:fits}).  We note that using a blackbody to model the soft excess component is not strictly physical, but provides a good representation of the spectral signatures and allows a simple comparison with other studies.  We only search for this component in sources with X-ray spectra containing at least 100 counts.  Using the F-test at 99\% significance, we detect this component in 60 of 680 sources.  Accounting for spurious detections, we find evidence for a soft excess component in 7.8\% of the sample.  This fraction agrees with previous studies of type 1 AGN \citep{mateos10, mateos05a}.  We note again that this is a lower limit of the percentage of sources we expect to have a soft excess component; this value could be as high at 80\%. 

In the automatic fitting process, \xspec is left to fit the po+bb model, which has 5 free parameters, with few constraints.  In 7 cases a flat power law and a large blackbody component gives a lower value of $\chi\sp{2}$, despite this fit being less physically realistic.  These sources were re-fit over the 0.5-12.0 keV detector frame energy range with the po+bb model, keeping the \gmm value fixed to that obtained from a fit to the data over the 2-10 keV rest frame energy range using just a simple power law.  \gmm and kT values more consistent with the other sources are then obtained.  An F-test between the manual fits and the simple power law fits is not appropriate since the new fit utilizes a result already obtained from the data.  It is therefore assumed that the blackbody component is still statistically required in the fit, based on the original classification.

\begin{figure}
  \centering
  \includegraphics[width=0.5\textwidth]{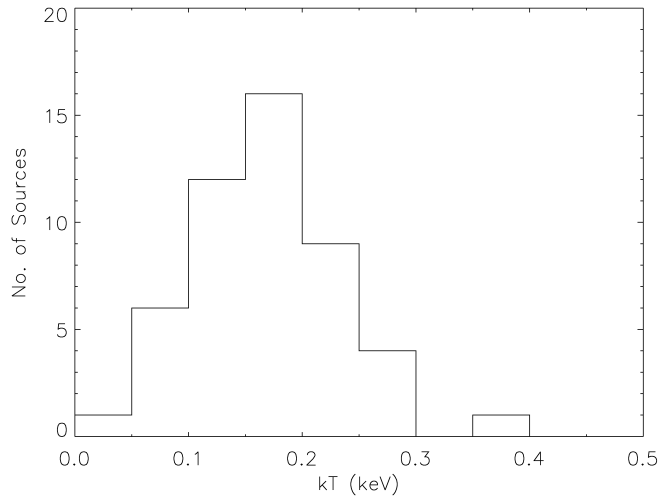}
  \caption{The distribution of kT values for the 56 sources in the sample that require the additional soft excess component and whose best fit provides an acceptable fit to the data.}
  \label{fig:kt_dist}
\end{figure}

The mean temperature of the soft excess component when fitted with a blackbody is $\left<\textrm{kT}\right>=0.17 \pm 0.09$ keV, obtained from modelling the distribution with a single Gaussian.  This is slightly higher than results in the literature \citep{mateos10,winter09,mateos05b,gierlinskidone04}, although consistent within errors.  The range in temperatures is observed to be rather small, with 68\% of the sources having kT values in the range 0.1-0.25 keV.  The full distribution of kT values can be seen in Figure~\ref{fig:kt_dist}.  

\begin{table*}
\begin{minipage}{170mm}
\centering
  \caption{The Spearman rank correlation coefficients and significances for correlations between the characteristic blackbody temperature used to model the soft excess, kT, and the parameters $z$, $L\sb{X}$, $L\sb{bb}$ and $M\sb{BH}$.  The best-fitting linear trends were determined by \chisq minimization.}
  \label{table:kt_corr}
     \centering
     \begin{tabular}{lll}
     \hline 
     Correlated Properties   & Strength of Correlation                               & Best-fitting Linear Trend          \\
     \hline
     kT and $z$              & $2.6\sigma$, $\rho=0.37$, prob=0.009                  & $\textrm{log}(kT)=(0.61 \pm 0.01)z-(1.059 \pm 0.004)$                    \\
     kT and \lumx            & $4.4\sigma$, $\rho=0.59$, prob$=1\times10\sp{-5}$     & $\textrm{log}(kT)=(0.08 \pm 0.01)\textrm{log}(L\sb{X})-(4.3 \pm 0.2)$    \\ 
     kT and $L\sb{bb}$       & $2.2\sigma$, $\rho=0.33$, prob=0.03                   & $\textrm{log}(kT)=(0.15 \pm 0.01)\textrm{log}(L\sb{bb})-(7.4 \pm 0.5)$   \\
     kT and $M\sb{BH}$       & $3.0\sigma$, $\rho=0.57$, prob=0.003                  & $\textrm{log}(kT)=(0.18 \pm 0.01)\textrm{log}(M\sb{BH})-(2.3 \pm 0.1)$   \\
     \hline
  \end{tabular}
\end{minipage}
\end{table*}

\begin{figure*}
  \centering
  \begin{tabular}{cc}
       \includegraphics[width=0.5\textwidth]{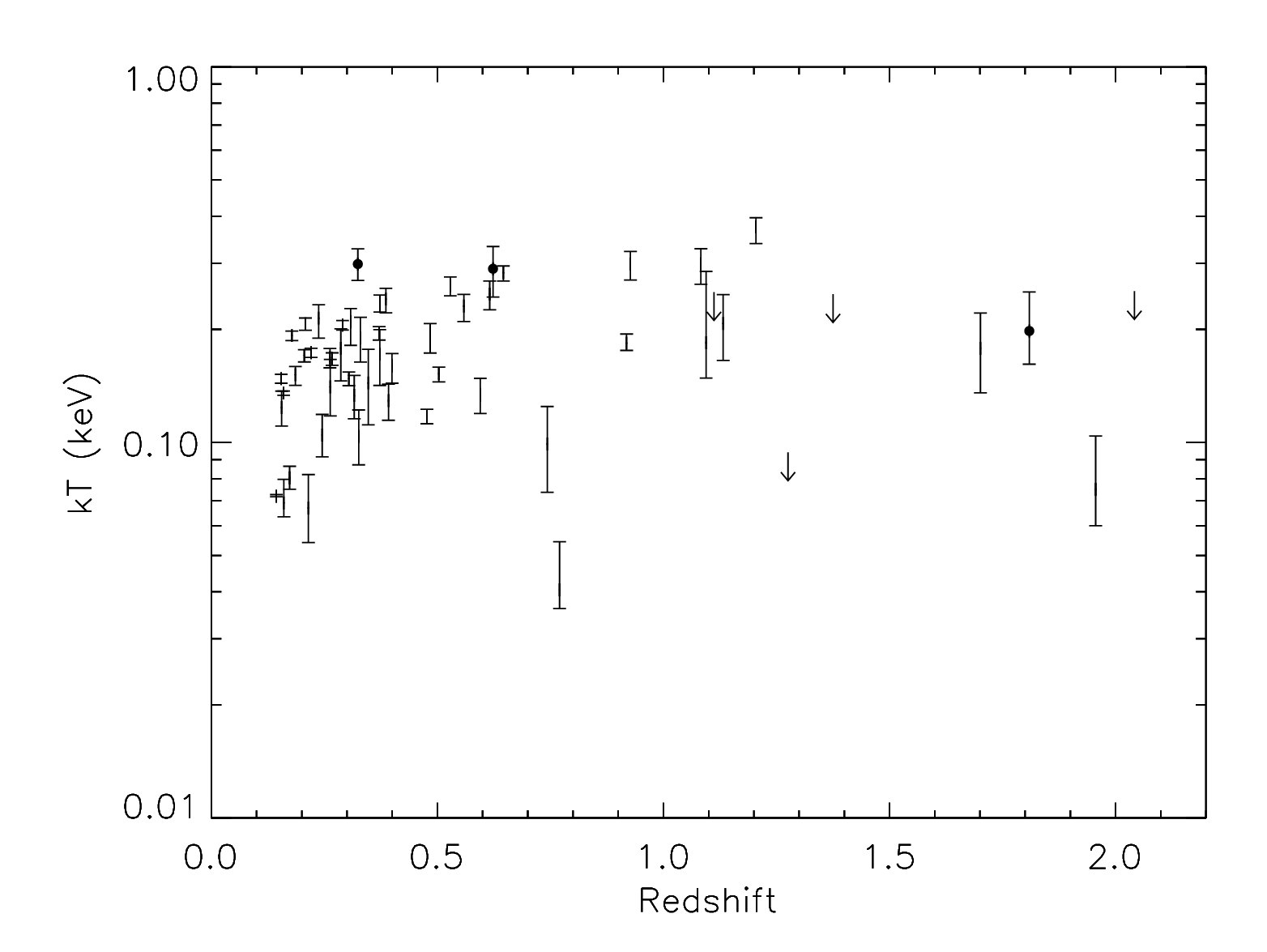}&
       \includegraphics[width=0.5\textwidth]{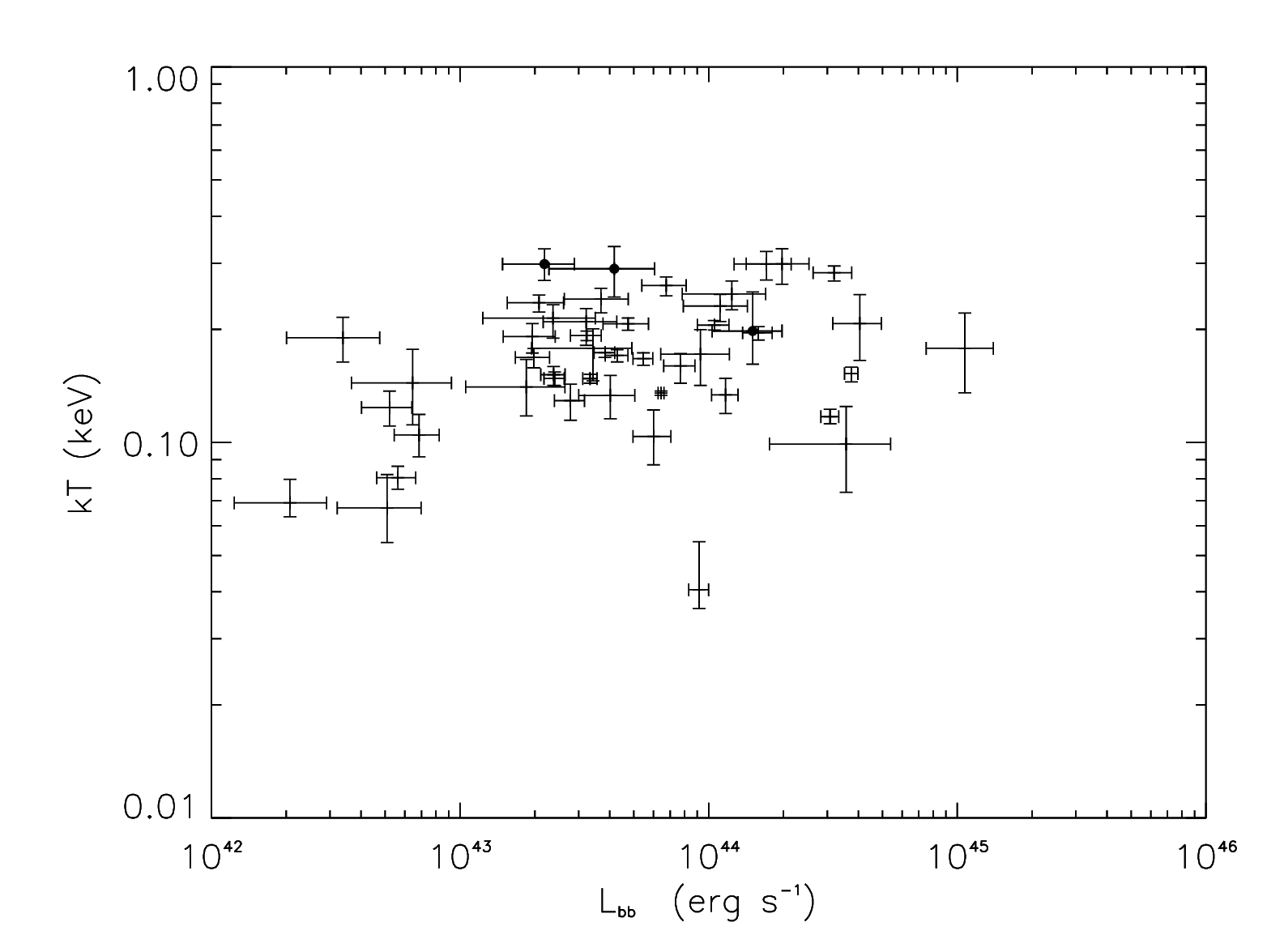}\\
       \includegraphics[width=0.5\textwidth]{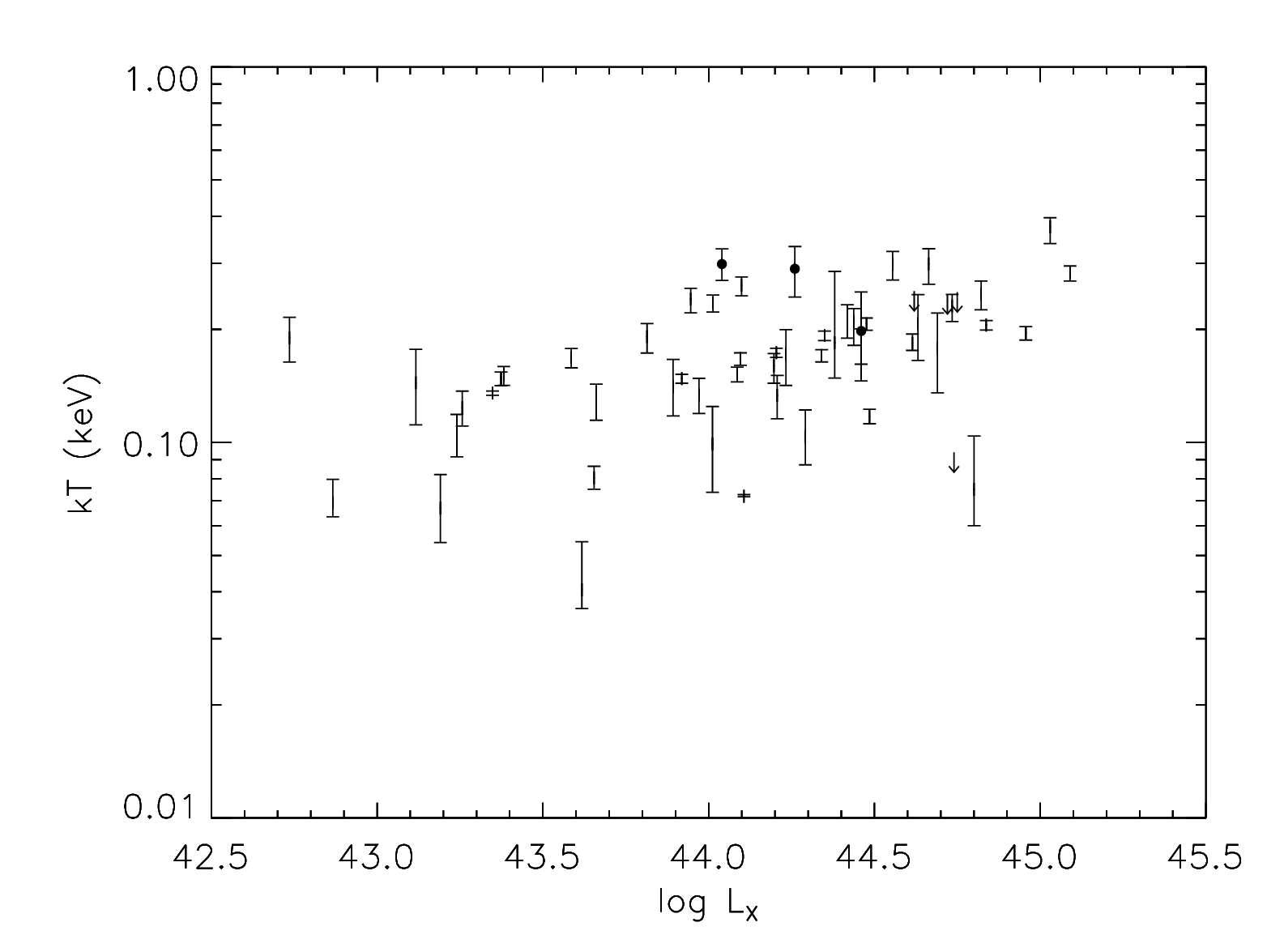}&
       \includegraphics[width=0.5\textwidth]{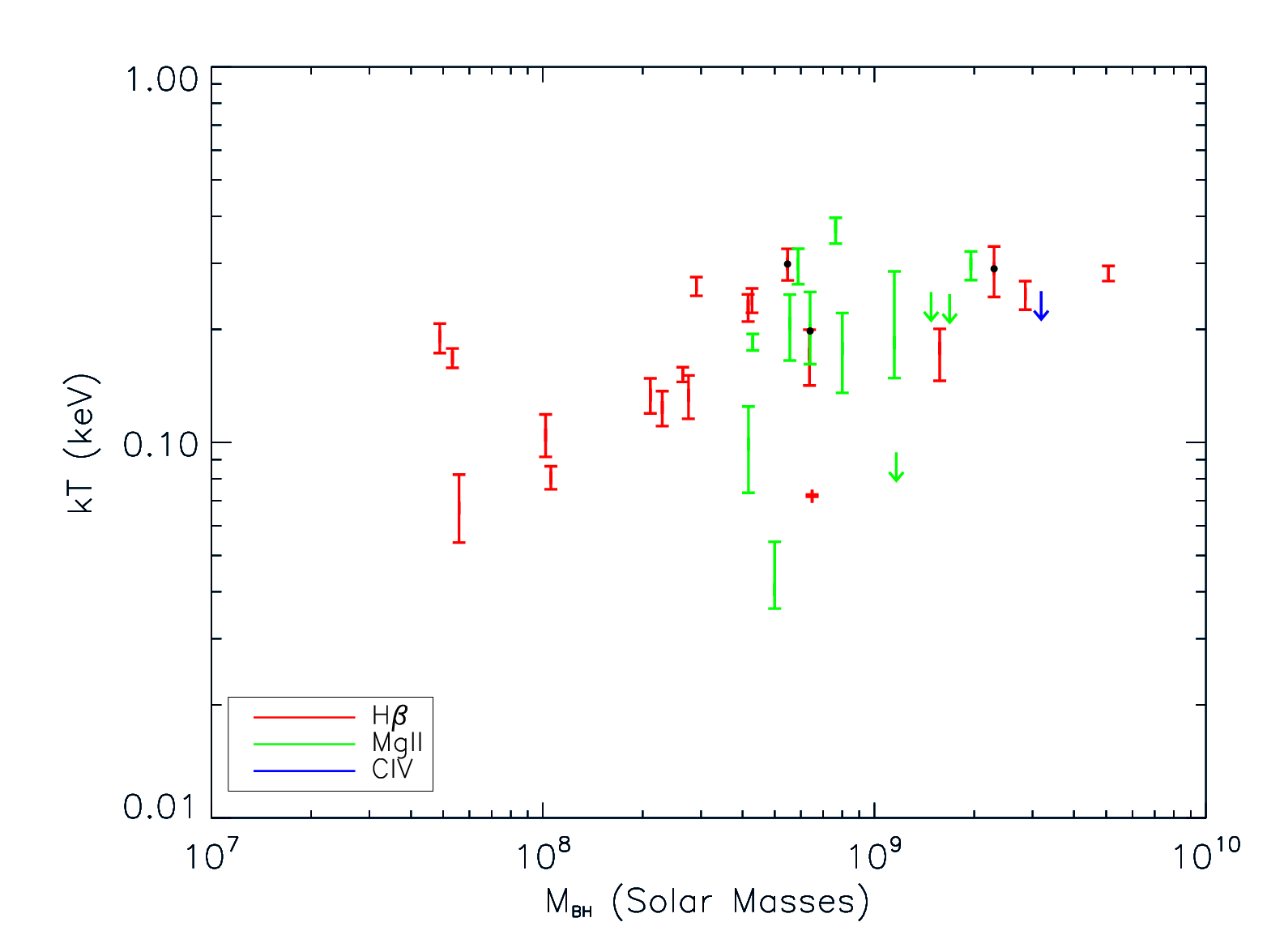}
  \end{tabular}
  \caption{The relations between the characteristic soft excess temperature, kT, and redshift (top left), hard X-ray luminosity (bottom left), blackbody luminosity (top right) and black hole mass (bottom left).  The error bars correspond to the 68\% confidence range.  Only sources for which H0$>$1\% are included on these plots.  Sources indicated by the filled black circles or upper limits are the ones fit with a manual po+bb model as mentioned in the text.  These sources are not included in the determination of any correlations or linear trends, however they are shown to be consistent with the rest of the sample.}
  \label{fig:kt_trends}
\end{figure*}

We consider possible correlations between the soft excess temperature $\textrm{log(kT)}$, and redshift, hard X-ray luminosity, the luminosity of the blackbody component ($L\sb{bb}$) and the black hole mass.  The correlation values and best-fitting linear trends can be found in Table~\ref{table:kt_corr}.  We find no significant correlation with $z$, which is consistent with previous results (e.g. \citealt{mateos10}).  Figure~\ref{fig:kt_trends} (top left) shows the presence of some soft excess components detected even in sources at $z>1$.

The standard explanation for the soft excess, and the motivation for modelling it with a blackbody is that it is thermal emission from the inner accretion disc.  The standard description predicts characteristic temperatures of the order kT $\sim$ 0.01 keV for a $10\sp{7} M\sb{\odot}$ black hole, as determined from $T\sb{disc} \propto M\sp{-1/4}(L/L\sb{Edd})\sp{1/4}$ \citep{shakurasunyaev73}.  We note that our average kT value is too high to be consistent with the standard disc origin.  This scenario also suggests that the temperatures should show variations with luminosity and black hole mass, with positive and negative linear trends, respectively.  

We find a significant correlation with \lumx (see Figure~\ref{fig:kt_trends} bottom left), but no significant correlation with $L\sb{bb}$.  A $3\sigma$ correlation is found between kT and $M\sb{BH}$\footnote{For 29 of the sources that required a blackbody component in the spectral fit, we also have a black hole mass estimate from the \citet{shen08} catalogue (See Section~\ref{section:black_hole} for details)}, but the best-fitting linear trend shows a positive, rather than negative gradient as would be expected.  This result disagrees with that of \citet{bianchi09a} who report the temperature of the blackbody component used to model the soft excess to be unrelated to the mass of the black hole. 

Previous studies have reported no trend of kT with luminosity \citep{bianchi09a, winter09}.  This problem was first highlighted by \citet{gierlinskidone04} and \citet{crummy06}, which led to the development of new theories, moving away from the accretion disc origin.  They invoke the idea that since the soft excesses display a narrow range of kT values, the physical process behind their origin is likely to be one of constant energy, such as atomic transitions.  At $\sim 0.7 $ keV, lines and edges of ionized O VII and O VIII produce a strong jump in opacity.  If these features were smeared by high velocities or gravitational redshifts, as found close to a black hole, a featureless soft excess would be produced.  Therefore the soft excess component could be an artefact of ionized absorption, likely from an outflowing disc wind \citep{gierlinskidone04}.  Alternatively, the partially ionized material could be out of the line of sight and seen via reflection.  In this scenario, the hard X-ray source needs to be at a small height above the disc so that light bending can illuminate the inner disc.  In this case the soft excess component is an artefact of relativistically blurred photoionized disc reflection \citep{ross05}.  

\begin{figure}
  \centering
  \includegraphics[width=0.48\textwidth]{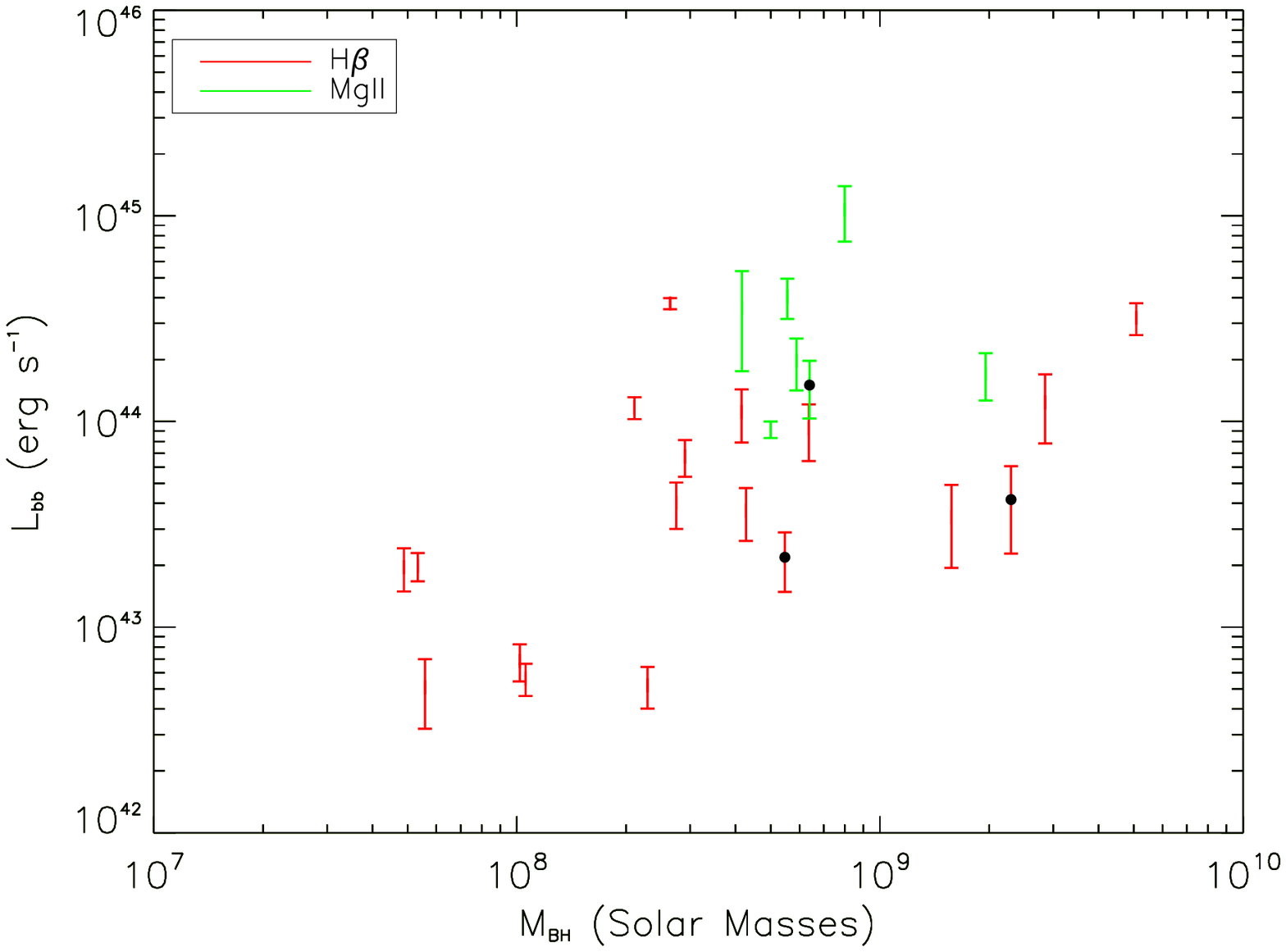}
  \caption{This figure includes the 25 sources best fit with the po+bb model, for which we also have a black hole mass estimate and a determination of $L\sb{bb}$.  The different colours correspond to the different emission line used in the mass determination (red - H$\beta$, green - MgII).  It shows the dependence of the black hole mass on the luminosity of the blackbody component, determined over the 0.5-10.0 keV range.  The errors shown on L$\sb{bb}$ correspond to the 68\% confidence range.  The black circles indicate the sources fit with the manual po+bb model.}
  \label{fig:mass_bb}
\end{figure}

In Figure~\ref{fig:mass_bb}, we plot black hole mass against the blackbody luminosity and find a $\sim 3\sigma$ correlation.  This suggests that regardless of the origin of the soft excess, its luminosity is likely related to the mass of the black hole.

\begin{figure}
  \centering
  \includegraphics[width=0.48\textwidth]{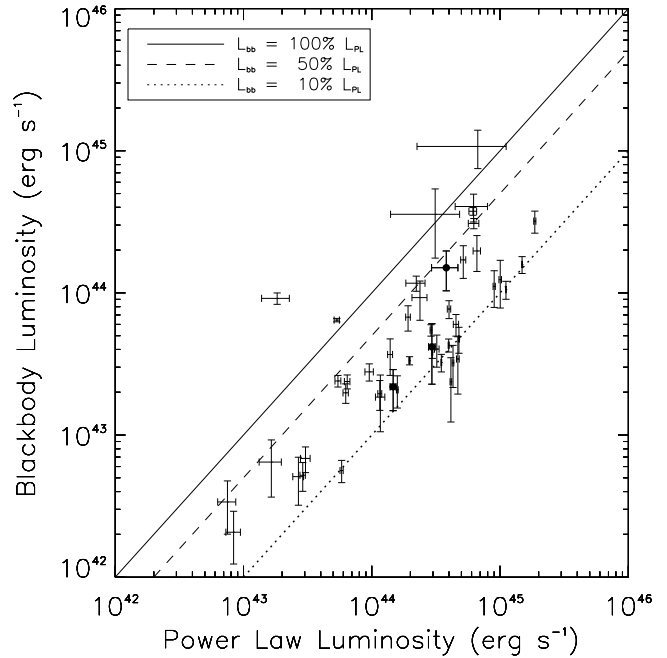}
  \caption{This plot compares the luminosity included in each of the separate components of the po+bb fits, calculated over the 0.5-10.0 keV range.  The linear trendline is $\textrm{log} L\sb{bb}=(0.78 \pm 0.02)\textrm{log} L\sb{PL}+(9.2 \pm 1.0)$.  The errors shown in these plots correspond to the 68\% confidence range.  The black circles indicate the sources fit with the manual po+bb model and are consistent with the rest of the sources.  The lines indicate the locations of $L\sb{bb}=L\sb{PL}$ (solid), $L\sb{bb}=50\%L\sb{PL}$ (dashed) and $L\sb{bb}=10\%L\sb{PL}$ (dotted) but do not represent any particular fit to the data.}
  \label{fig:Lbb}
\end{figure}

Figure~\ref{fig:Lbb} shows a strong ($\sim8\sigma$) correlation between the luminosity of the blackbody component and the luminosity of the power law component with a best-fitting trend of $\textrm{log} L\sb{bb}=(0.78 \pm 0.02)\textrm{log} L\sb{PL}+(9.2 \pm 1.0)$.  This indicates a link between the energy production in these 2 components and supports the idea of inverse Compton scattering where the low energy photons from the disk are up-scattered to X-ray energies creating the power law emission.  However it could also imply a reverse process in which the power law emission is re-processed into the lower energy soft excess.  

This correlation has been previously reported in \citet{winter09}, where they find $L\sb{PL}=(0.79 \pm 0.14)L\sb{bb} + (9.34 \pm 6.04)$.  A similar correlation is seen in \citet{mateos10} where the power law luminosity over 2-10 keV against the blackbody luminosity over 0.5-2 keV is plotted.  The best-fitting linear function is found to be $\textrm{log} L\sb{bb}=(1.2 \pm 0.1)\textrm{log} L\sb{PL}+(9.2 \pm 5.3)$.  It is suggested that the lack of sources in the bottom right of the plot is a selection effect due to the fact that a low luminosity blackbody component will be difficult to detect over a high luminosity power law component and the lack of sources in the top left of the plot is due to an apparent upper limit to the soft excesses allowed in AGN.  For comparison, the linear trend best fit to our data when the errors are not considered is $\textrm{log} L\sb{bb}=(0.80 \pm 0.09)\textrm{log} L\sb{PL}+(8.3 \pm 4.1)$.  These linear relationships each with gradients close to 1 imply that the fraction of converted emission is constant for sources of all luminosities.  However all 3 data sets show that in the luminosity range $10\sp{42} - 10\sp{46} \textrm{erg s}\sp{-1}$, the luminosity in the power law is greater than that in the blackbody component.

\subsection{Accretion Properties}
\label{section:black_hole}

Since the X-ray emission is believed to originate from the accretion process, relationships between the X-ray spectral properties and the physical properties of the central engine are expected.  In this section we consider correlations between both the broad-band continuum and the soft excess properties with Eddington ratio.  We also investigate any differences in Eddington ratio between the radio loud and radio quiet sub-samples.

Black hole masses can be determined for type 1 AGN by assuming that the broad line region is gravitationally bound to the central black hole potential and that the broad line clouds are virialized.  By measuring the width of broad emission lines produced by these clouds their velocity ($V\sb{g}$) can be estimated.  The radius of the broad line region ($R\sb{BLR}$) can be estimated from a linear correlation with the monochromatic optical luminosity.  The mass of the black hole can then be determined from the equation, $M\sb{BH} = R\sb{BLR} V\sp{2}\sb{g} G\sp{-1}$ \citep{kaspi00}.  

Here we use the \citet{shen08} catalogue of black hole masses estimated using this technique, which includes $\sim 58,600$ quasars from the SDSS DR5 quasar catalogue.  The mass is estimated using a different broad line, depending upon the redshift of the source.  \hb is used for $z<0.7$, MgII for $0.7<z<1.9$ and CIV for $z>1.9$.  The catalogue also includes bolometric luminosities estimated from optical monochromatic luminosities (at $5100\textrm{\AA}$ for H$\beta$, $3000\textrm{\AA}$ for MgII and $1350\textrm{\AA}$ for CIV), derived using bolometric correction factors from \citet{richards06}.  Of the 734 sources (with $\textrm{H0}>1\%$), 79\% have a black hole mass estimate available in the \citet{shen08} catalogue.  This gives us a sample of 581 sources.

The Eddington luminosity is calculated using Equation~\ref{equation:eddington}.  The bolometric luminosity is calculated from the observed X-ray luminosity in the 2-10 keV band, $L\sb{X}$, using the luminosity dependent bolometric correction of \citet{marconi04} described by Equation~\ref{equation:marconi} where $\mathcal{L}=\textrm{log} L\sb{bol} - 12$ and $L\sb{bol}$ is in units of $L\sb{\odot}$.  The Eddington ratio is then defined as $L\sb{bol}/L\sb{Edd}$.

\begin{equation}
\label{equation:eddington}
L\sb{Edd}=\frac{4\pi G M m\sb{p} c}{\sigma\sb{c}} \sim 1.3\times10^{38}\frac{M}{M\sb{\odot}} \rm{erg\,s}^{-1}
\end{equation}     

\begin{equation}
\label{equation:marconi}
  \textrm{log}(L\sb{bol}/L\sb{X}) = 1.54 + 0.24 \mathcal{L} + 0.012 \mathcal{L}\sp{2} - 0.0015 \mathcal{L}\sp{3}
\end{equation}

\begin{figure}
  \centering
  \includegraphics[width=0.5\textwidth]{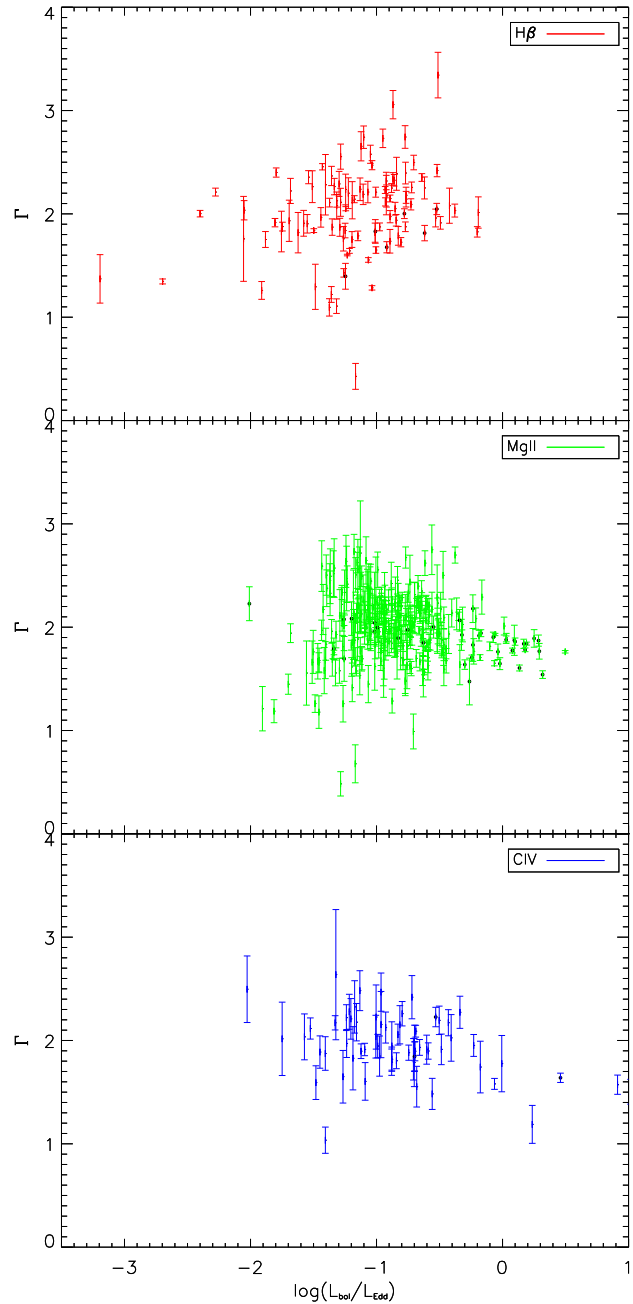}
  \caption{This figure shows the relationship between \gmm and the Eddington ratio, where the latter is based on mass estimates obtained using different broad emission lines.  Top: 117 sources for which the \hb line was used.  Middle: 385 sources for which the MgII line was used.  Bottom: 79 sources for which the CIV line was used.  The error bars correspond to 68\% errors.  Sources indicated with the black dots are RLQ.}
  \label{fig:edd}
\end{figure}

In Figure~\ref{fig:edd} we plot the relationship between \gmm and the Eddington ratio, separated depending on which broad line was used to estimate the black hole mass.  We find a positive correlation for the \hb data (Spearman Rank correlation coefficient, $\rho=0.26$, $2.8\sigma$), no correlation for the MgII data and a negative correlation for the CIV data ($\rho=-0.32$, $2.9\sigma$), however the reliability of the CIV line as a mass estimator has been widely questioned (e.g. \citealt{shen08}).  This differs slightly with the results of \citet{risaliti09} who find a highly significant, positive correlation with estimates based on the H$\beta$ line, a significant, although weaker positive correlation with MgII and no significant correlation with CIV.  We note that in their analysis they use the bolometric luminosity given by \citet{shen08} and use the power law slope determined in the 2-10 keV rest frame band which may explain these differences.  Previous studies of PG Quasars with \xmm have reported a positive correlation between \gmm and Eddington rate \citep{porquet04, piconcelli05} and \citet{shemmer06} show that this is the dominant correlation, as opposed to the known relation between accretion rate and line width.  We note that whilst we use a luminosity dependent bolometric correction to obtain our bolometric luminosity estimates, recent works have indicated that the bolometric correction factor may be dependent on the Eddington ratio itself \citep{vasudevan07}.  

The previous section indicated a strong relationship between the luminosities in the blackbody component and the power law component.  Since the properties of the power law component may correlate with the Eddington ratio, we also investigate the dependence of the blackbody properties on Eddington ratio.  We consider the dependence between kT, blackbody luminosity and black hole mass on the Eddington ratio.  No significant correlations are found ($\textrm{kT}=0.8\sigma$, $L\sb{bb}=2.4\sigma$, $M\sb{BH}=0.3\sigma$).

\begin{figure*}
  \centering
  \begin{tabular}{cc}
   \includegraphics[width=0.42\textwidth]{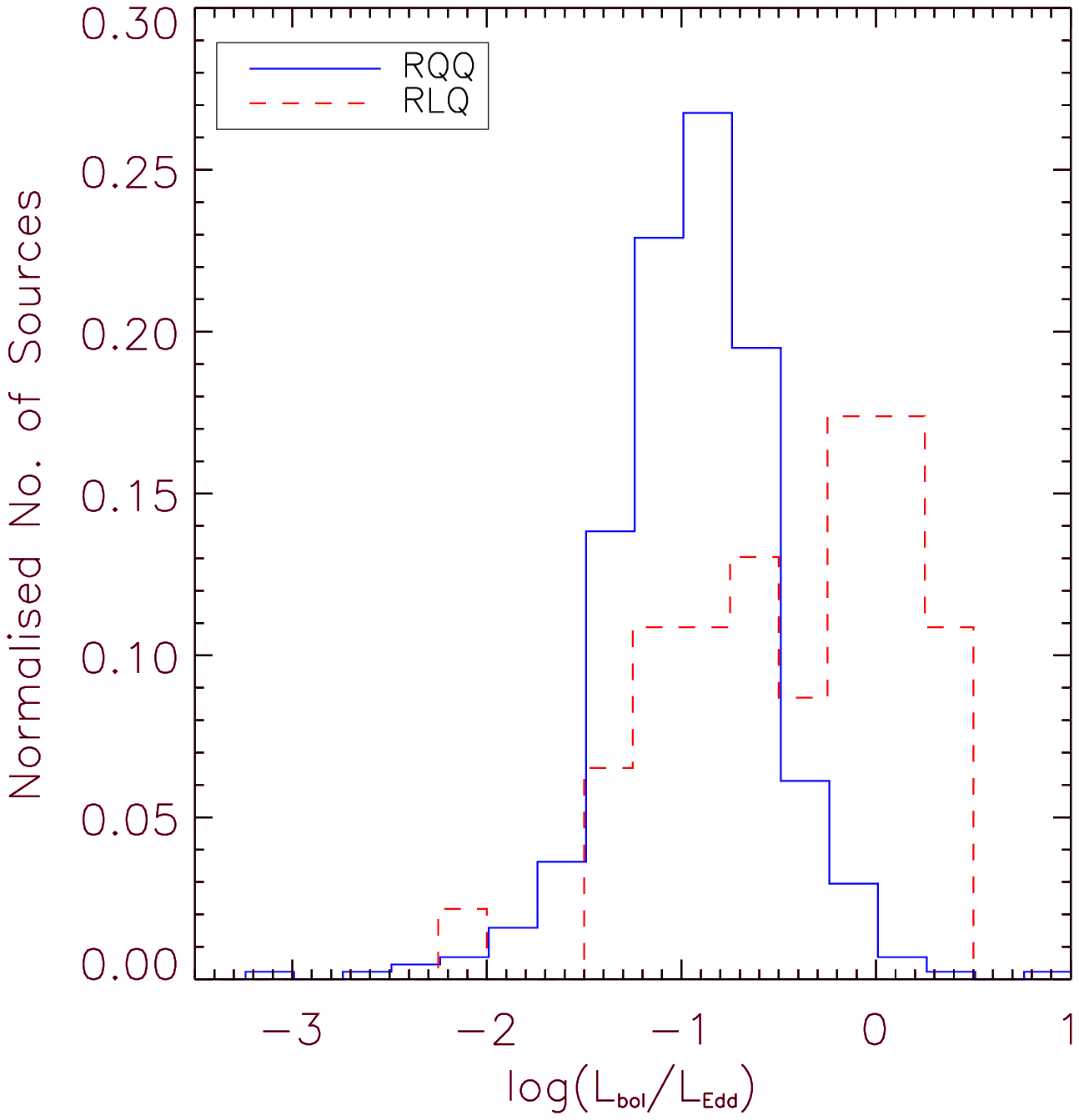} &
   \includegraphics[width=0.42\textwidth]{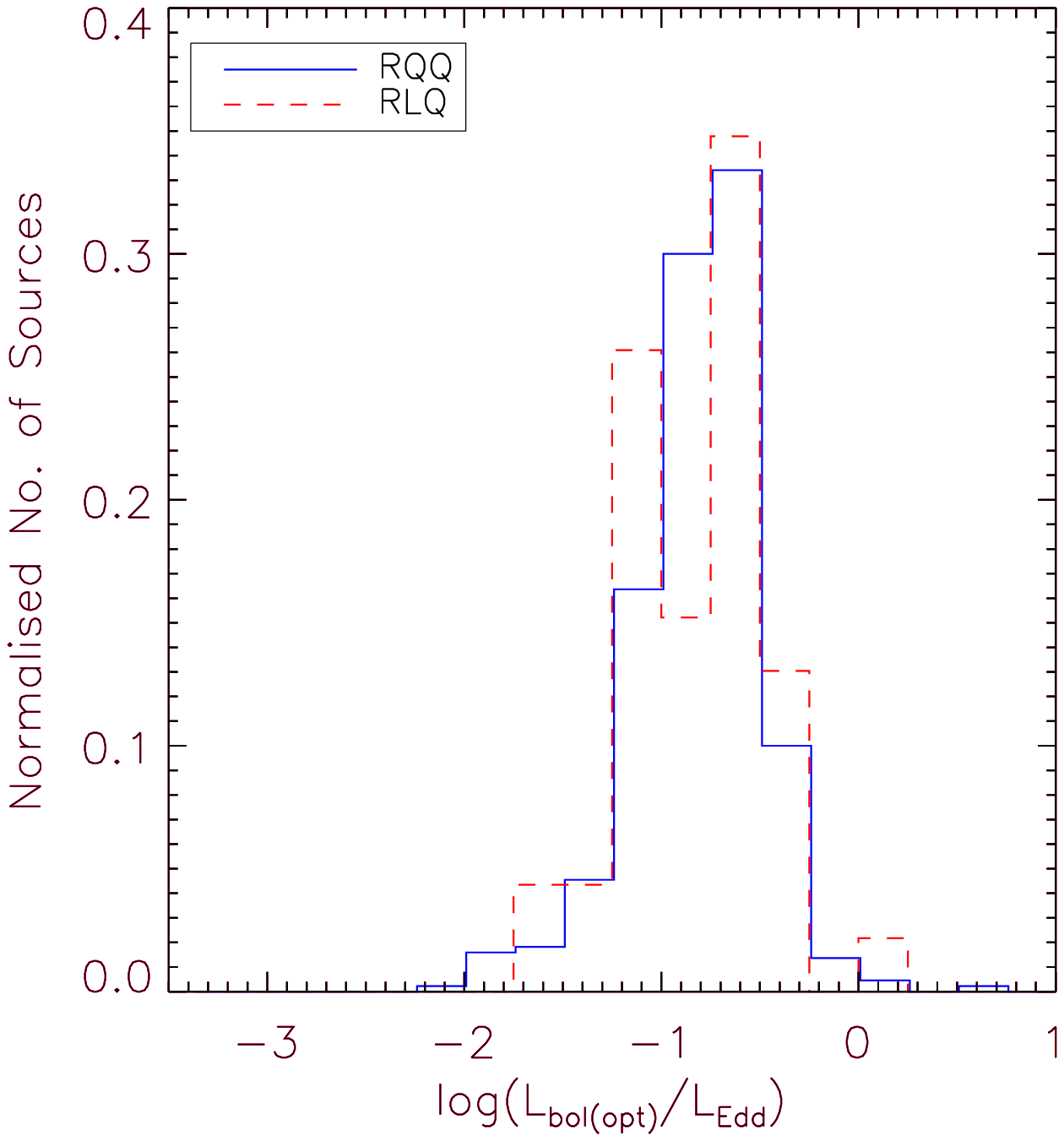}\\
   \vspace{3mm}
   \includegraphics[width=0.48\textwidth]{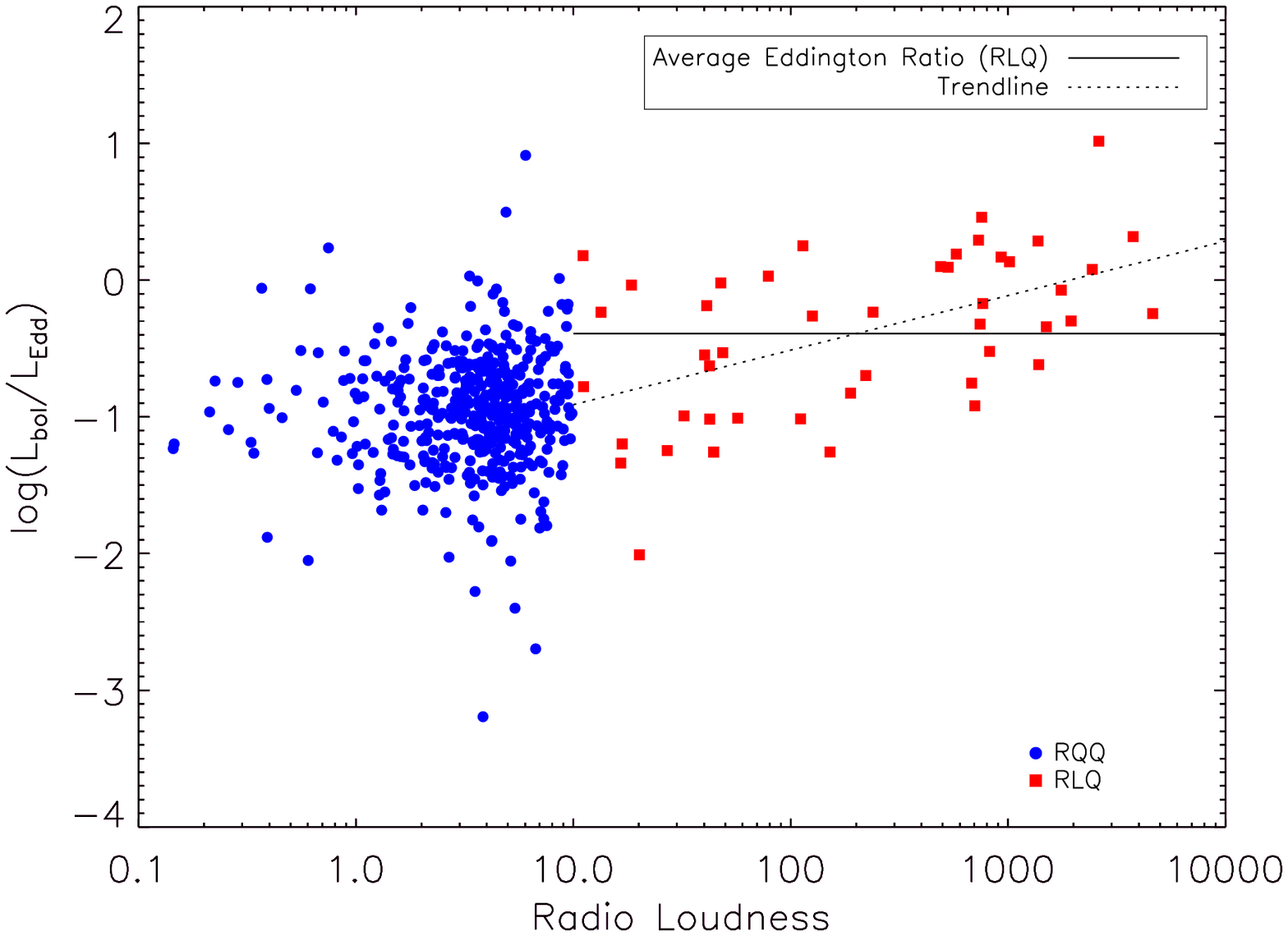} &
   \includegraphics[width=0.48\textwidth]{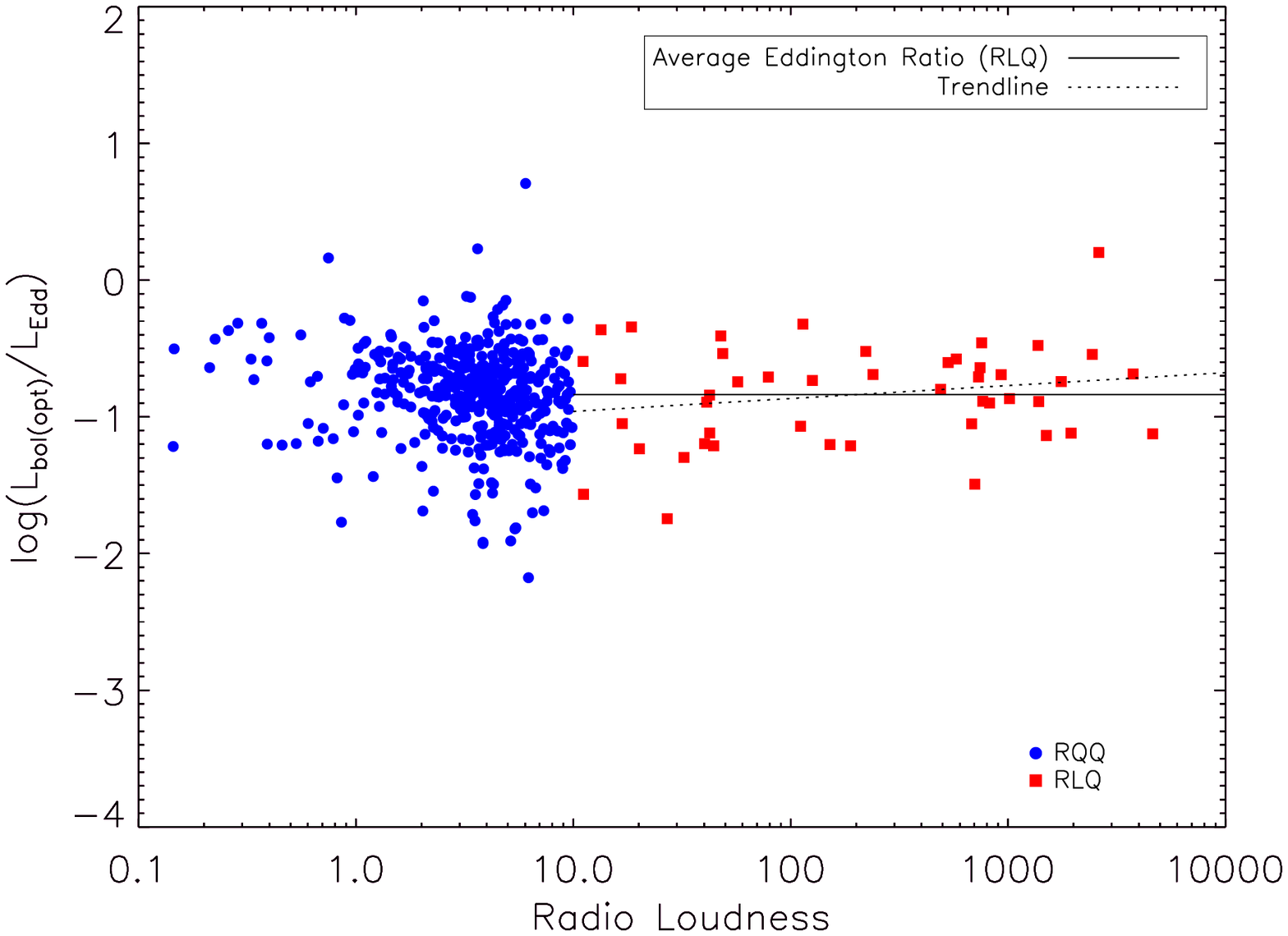}\\
   \end{tabular}
  \caption{The top figures show the Eddington ratio distributions of the confirmed radio quiet (solid blue) and radio loud (dashed red) sources. The bottom figures show the trend of $R\sb{L}$ with the Eddington ratio with the best-fitting trend to the RLQ shown by the dashed line.  The figures on the left use Eddington ratios defined from the observed 2-10 keV X-ray luminosity and the luminosity dependent bolometric correction of \citet{marconi04}.  The figures on the right use Eddington ratios defined from monochromatic luminosities in the optical and UV by \citet{shen08}, and bolometric correction factors from \citet{richards06}.} 
  \label{fig:edd_radio}
\end{figure*}

In order to investigate a possible link between radio jet production and accretion rate, we have compared the distribution of Eddington ratios for the confirmed RLQ and RQQ sub-samples and investigated a possible trend between radio loudness and Eddington ratio.  We consider both an X-ray derived Eddington ratio, (as previously described), and an optically derived Eddington ratio using the bolometric luminosities derived by \citet{shen08} (also described above).  The distributions of Eddington ratios are plotted in Figure~\ref{fig:edd_radio} (top) and the KS significances are listed in Table~\ref{table:radio_eddington_corr}.  Figure~\ref{fig:edd_radio} (bottom) shows the correlation between radio loudness, $R\sb{L}$ and Eddington ratio, with the best-fitting linear trends listed in Table~\ref{table:radio_eddington_corr}.

\begin{table*}
\begin{minipage}{170mm}
\centering
  \caption{Listed in the first column are the KS significance levels found when comparing the distribution of Eddington ratios (derived from both X-ray and optical bolometric luminosities) for RQ and RL sub-samples.  Also listed are the best-fitting linear trends between radio loudness, $R\sb{L}$, and Eddington Ratio, for the combined RQ and RL populations, and just the RLQ on their own.}
  \label{table:radio_eddington_corr}
     \centering
     \begin{tabular}{l|l|ll}
     \hline 
               & KS significance       & Sources included in trend  & Best-fitting Linear Trend                                                      \\
     \hline
     X-ray     & $2\times 10\sp{-10}$  & RLQ \& RQQ       &  log($L\sb{bol}/L\sb{Edd}$) = $(0.19 \pm 0.03)\textrm{log} R\sb{L} - (1.03 \pm 0.03)$    \\
               &                       & RLQ              &  log($L\sb{bol}/L\sb{Edd}$) = $(0.4 \pm 0.1)\textrm{log} R\sb{L} - (1.3 \pm 0.2)$        \\
     Optical   & $0.50$                & RLQ \& RQQ       &  log($L\sb{bol(opt)}/L\sb{Edd}$) = $(-0.02 \pm 0.02)\textrm{log} R\sb{L} - (0.81 \pm 0.02)$   \\
               &                       & RLQ              &  log($L\sb{bol(opt)}/L\sb{Edd}$) = $(0.09 \pm 0.07)\textrm{log} R\sb{L} - (1.05 \pm 0.17)$    \\
     \hline
  \end{tabular}
\end{minipage}
\end{table*}

We find that the distribution of X-ray derived Eddington ratios is significantly different between the RQQ and RLQ sub-samples, with RLQ having higher Eddington ratios.  However, when the optically derived Eddington ratios are compared, no significant difference is found.  Similarly, a significant trend towards higher X-ray derived Eddington ratios in sources with a higher radio loudness is suggested.  This trend is not found when optically derived Eddington ratios are used.

We suggest that the differences found between using X-ray derived and optically derived Eddington ratios may be an artefact of the additional X-ray component suggested to be present in radio loud quasars (see Section~\ref{section:radio}).  Whereas the optical emission of RQQs and RLQs is similar (the distributions of $M\sb{i}$ were found to not be significantly different), the X-ray luminosity of the radio loud sources was found to be higher, which results in relatively higher X-ray derived Eddington ratios for these sources.


\section{Discussion}
\label{section:discussion}
Much of the information gained to date on the X-ray emission mechanisms of AGN and their interplay with the surrounding environment has come from detailed studies of bright AGN, which reveal significant spectral complexities (e.g. \citealt{pounds04}).  The nature of our sample here means that despite dealing with lower count spectra which allow only relatively simple models to be fit, we have a sufficiently large sample with which to investigate source-to-source variations and the range in spectral properties. 

At the simplest level the sample is well described by a simple power law model. There is significant intrinsic dispersion in the power law indices describing the broad-band X-ray continuum, with a small fraction of objects which are either extremely flat or extremely steep. We emphasize that for a number of these extreme objects we have sufficient counts to rule out effects caused by the non-inclusion of additional spectral complexity.  The distribution is reasonably symmetric about the mean, suggesting that the underlying electron acceleration mechanism and/or Comptonisation are subject to a random scatter, as opposed to a limiting cascade mechanism \citep{svensson94} which might be expected to produce a larger asymetry in the distribution of values.

Despite evidence that nearby AGN display rapidly variable spectra (e.g. \citealt{vaughan11}), we assume constant X-ray spectra for our sources.  However, a Principal Component Analysis (PCA; \citealt{PCA}) on a long \xmm observation of MCG-6-30-15, found that 96\% of the variability is due to the continuum component, the spectrum of which is well fit with a simple power law of photon index \gmm= $2.20 \pm 0.05$.  The spectral variations are thought to be caused by the normalization of this power law component changing relative to a constant component, rather than changes in the underlying power law index \citep{vaughan04}.  The value of this is found to be similar to the average \gmm value of our sample.

The concept of Black Hole Unification says that since AGN and Galactic Black Hole Binaries (GBHBs) are both examples of accreting black holes, they should be expected to display similar characteristics, in which timescales simply scale with mass.  In particular, GBHBs exist in different spectral states corresponding to different accretion rates with transitions between states occuring on the order of weeks.  For AGN this timescale is $\sim 10\sp{5}$ years, meaning that such transitions are not observed for individual AGN.  An analogy is made between the spectral states of GBHBs and the different types of AGN with some RLQ perhaps being objects transitioning to the high/soft state via the jet line \citep{fender06}.  We expect the majority of objects in our sample to be in the high/soft state due to selection biases.

The most significant factor to correlate with spectral index is the intrinsic X-ray luminosity of the object.  This finding is similar to that of \citet{young09} who report no correlation with redshift, but a strong correlation with luminosity.  They report strong correlations of \gmm with rest-frame monochromatic luminosities at 0.7, 2.0, 10.0 and 20.0 keV, with the strength of the correlation increasing towards higher energies.  However, since the monochromatic luminosity determination depends directly on the \gmm value, a correlation between the two quantities is expected.  As a flat source has more emission at higher energies than a steep source, it is the flatter sources that will have the higher monochromatic luminosities at these energies.  This effect will decrease between $l\sb{20 keV}$ and $l\sb{2 keV}$, until a pivot point at which the steeper sources start to have more emission at the energy at which the monochromatic luminosity is calculated, than the flat sources.

Calculating the Eddington ratio requires an estimate of the bolometric luminosity of the source.  This can be inferred from either the observed X-ray or optical luminosity and we obtain different estimates depending on which waveband is used.  In particular we show that RQQ and RLQ have similar optically derived Eddington ratios but higher X-ray derived Eddington ratios, likely due to the generally higher 2-10 keV X-ray luminosities seen in RLQ, on which the ratio is based.  Whilst we note that an inherent bias due to the RLQ is present, we see a trend for an increased Eddington ratio with increasing radio loudness, which might argue for models in which jet formation and hence radio emission depend on the accretion mode/Eddington ratio in analogy with galactic stellar mass black holes.  We also see some evidence for the correlation of spectral index with radio loudness.  In particular we show that the RLQs have on average flatter power law slopes which may be related to the synchrotron emission from a jet whose importance depends on the radio luminosity.  Therefore for objects with higher Eddington ratios we expect a higher radio loudness and from this a flatter power law spectrum.  There is however conflicting evidence for a correlation of the power law index with the Eddington ratio.  

In agreement with previous works \citep{porquet04,piconcelli05,shemmer06,risaliti09} we find a, positive correlation between \gmm and the Eddington ratio.  However, this is only true when the black hole mass estimates are determined using the width of the broad \hb line; if the MgII line is used, no correlation is seen, and the correlation appears to switch direction when using the CIV line.  The virial method for determining black hole masses is calibrated using results from reverberation mapping studies.  These are only widely available for low redshift objects, giving a well calibrated method using the \hb broad line but a less reliable calibration for the MgII and CIV lines which are simply scaled to match the results from the \hb line.  \citet{shen08}, from whom the black hole mass estimates are taken, note that the estimates determined from the \hb and MgII lines tend to agree with each other, but the masses estimated from the CIV line do not.  The CIV lines can be severely affected by a disc wind component and therefore may be asymetric and blue shifted, leading to an overestimate of the width of the line and a FWHM which is not representative of the virial velocity of the broad line region.  However, some authors have reported consistency between black hole masses determined with the H$\beta$, MgII and CIV lines \citep{assef10,greene10}.  Whilst the different correlations we see may be due to a calibration effect, we cannot rule out the possibility of a hidden physical effect.  By necessity different broad lines are used for sources at different redshifts and the CIV line is used for sources at high redshifts and likely high X-ray luminosities.  Therefore the difference in these parameters may be the true driver behind the different trends we see with Eddington ratio. 

Over recent years there has been much debate about the origin of the soft excess detected below $\sim 0.5$ keV in AGN. Initially perceived as simply blackbody emission from the accretion disc \citep{turner89} the effective blackbody temperatures measured are significantly higher than accretion disc models allow. Alternative models have been proposed, including Compton upscattering of the disc emission \citep{kawaguchi01}, reflection of the power law emission from the disc \citep{crummy06}, and absorption of the power law emission by ionised material \citep{gierlinskidone04}. Our results, based on fitting this component with a blackbody, argue against, but do not rule out, these absorption models, for which a correlation between the apparent temperature of the blackbody and fractional luminosity of the apparent excess might be expected.  We find a strong correlation between the luminosities of the soft excess and the power law components, with the soft excess luminosities being typically 10-50\% that of the power law.  This is easier to explain with an emission process such as Compton upscattering.


\section{Conclusions}
\label{section:conclusion}
In this paper the X-ray spectra of a sample of 761 type 1 AGN from a cross-correlation of the SDSS DR5 quasar catalogue and 2XMMi have been analysed.  The main conclusions from this analysis of the largest sample of AGN X-ray spectra are summarized below:

\begin{enumerate}

\item When the distribution of best-fitting power law slopes is modelled with a single Gaussian as done in previous works, the best-fitting values are found to be \wgmm$= 1.99 \pm 0.01$ with an dispersion of \sigwgmm $= 0.30 \pm 0.01$.  The typical 68\% error on a source is $\Delta\Gamma = 0.13$, which is much lower than the intrinsic dispersion of values.  The distribution is better fit with a combination of 2 Gaussians due to the presence of extreme sources with flat or steep \gmm values. 
\item A marginal trend ($\sim 3\sigma$) for flatter \gmm values in higher redshift sources is found, but only when just those sources best fit with a simple power law model are used.  This apparent flattening does not appear to be due to the presence of a reflection component in the spectrum at higher energies.
\item A significant trend ($\sim 5\sigma$) for flatter \gmm values in higher X-ray luminosity sources is found.
\item The radio loud quasars are found to have higher X-ray luminosities compared to the radio quiet quasars.  We suggest this may be due to an additional X-ray component, perhaps related to a jet, which also causes RLQ to show flatter \gmm values and higher Eddington ratios, if $L\sb{bol}$ is calculated from the X-ray luminosity. 
\item We find evidence for intrinsic neutral absorption in 3.4\% of our sample.  However, our simulations suggest that the true fraction of absorbed type 1 sources could be as high as $\sim 25\%$.  The levels of absorption ($N\sb{H} = 10\sp{21}-10\sp{23} \textrm{ cm}\sp{-2}$) are typically much higher than the optical extinction of the broad-lines and those expected for type 1 objects.   No trend is found between $N\sb{H}\sp{intr}$ and redshift or X-ray luminosity. The fraction of absorbed sources does not appear to vary with luminosity or redshift.
\item We find evidence for a soft excess component in $\sim 8\%$ of our sample, but simulations suggest that up to 80\% of our sources may have a soft excess.  The average blackbody temperature used to model the soft excess is $\left<\textrm{kT}\right> = 0.17 \pm 0.09$ keV. This temperature is shown to correlate with hard luminosity.
\item A strong correlation ($\sim 8\sigma$) is found between the blackbody luminosity and power law luminosity for those sources best fit with the po+bb model with the soft excess luminosity being $\sim 10-50\%$ that of the power law.
\item A positive correlation ($\rho=0.26$, 2.8$\sigma$) of \gmm with the X-ray derived Eddington ratio is found when the broad \hb line is used in the determination of the black hole mass.  No such trend is found when the MgII line is used and a negative correlation ($\rho=-0.32$, 2.9$\sigma$) is found when the CIV line is used.  The blackbody properties kT, $L\sb{bb}$ and $M\sb{BH}$ are not found to correlate with Eddington ratio.
\end{enumerate}

\section*{Acknowledgments}
We thank the referee for useful comments which helped improve the paper.  We also thank Ken Pounds and Valentina Braito for comments and suggestions.  AES and SH acknowledge support from STFC studentships.  SM acknowledges financial support from the Ministerio de Ciencia e Innovación under project AYA2009-08059 and AYA2010-21490-C02-01.  DMA thanks the Royal Society, Leverhulme Trust and STFC for support.  This work was based on observations obtained with \textit{XMM-Newton}, an ESA science mission with instruments and contributions directly funded by ESA Member States and NASA.  Funding for the SDSS and SDSS-II has been provided by the Alfred P. Sloan Foundation, the Participating Institutions, the National Science Foundation, the U.S. Department of Energy, the National Aeronautics and Space Administration, the Japanese Monbukagakusho, the Max Planck Society, and the Higher Education Funding Council for England.

\bibliographystyle{mn2e}
\bibliography{mybib}

\bsp

\label{lastpage}

\end{document}